\documentclass[11pt]{amsart}
\usepackage[a4paper,margin=25mm]{geometry}
\usepackage[T1]{fontenc}
\usepackage{lmodern,amsmath,amssymb,amsthm,mathtools,mathrsfs,microtype,booktabs}
\usepackage[hidelinks]{hyperref}
\numberwithin{equation}{section}
\allowdisplaybreaks[2]
\newtheorem{theorem}{Theorem}[section]
\newtheorem{lemma}[theorem]{Lemma}
\newtheorem{proposition}[theorem]{Proposition}
\newtheorem{corollary}[theorem]{Corollary}
\theoremstyle{remark}

\DeclareMathOperator{\Ric}{Ric}
\DeclareMathOperator{\Scal}{Scal}
\DeclareMathOperator{\Vol}{Vol}
\DeclareMathOperator{\PU}{PU}
\DeclareMathOperator{\SU}{SU}
\DeclareMathOperator{\SO}{SO}

\DeclareMathOperator{\wt}{wt}
\DeclareMathOperator{\Tr}{Tr}
\DeclareMathOperator{\ad}{ad}
\newcommand{\E}{\mathbb E}
\newcommand{\Pp}{\mathbb P}
\newcommand{\HH}{\mathcal H}
\newcommand{\CC}{\mathcal C}
\newcommand{\MM}{\mathcal M}
\newcommand{\F}{\mathbb F}
\title[Einstein metrics and quantum control]{Einstein metrics and quantum control on symmetric tensor product quotients}
\author{Honghuai Fang}
\address{Institute for Theoretical Sciences, Westlake University, Hangzhou, China}
\email{fanghonghuai@westlake.edu.cn}
\date{}
\hypersetup{pdftitle={Einstein metrics and quantum control on symmetric tensor product quotients},pdfauthor={Honghuai Fang},pdfsubject={Einstein classification, invariant control distances, and two-body locality}}

\DeclareMathOperator{\vrad}{vrad}
\DeclareMathOperator{\Arg}{Arg}
\DeclareMathOperator{\Ent}{Ent}
\DeclareMathOperator{\diam}{diam}

\begin{document}
\begin{abstract}
Let $M_n=(\mathrm{PU}(2)^n\rtimes S_n)\backslash\mathrm{PU}(2^n)$ be the
space of projective $n$-qubit unitaries modulo local unitaries and
permutations of the qubits. We prove that for all sufficiently large $n$
the manifold $M_n$ carries exactly two $\mathrm{PU}(2^n)$-invariant Einstein
metrics up to scaling. Both Einstein metrics have uniformly bounded diameter after normalizing two-qubit costs, although approximating a Haar-random unitary requires exponentially many two-qubit gates. In contrast, we construct non-Einstein metrics with nonnegative Ricci curvature and exponentially growing typical control distances, and determine the sharp curvature threshold for this family.
\end{abstract}
\maketitle

\section{Introduction}
For a compact homogeneous space $G/H$, the Einstein condition on
$G$-invariant metrics is a finite system of algebraic equations in the
metric coefficients. Writing down this system is routine; finding all of
its positive solutions is not. The coefficients range over a noncompact
cone, and a solution found near a convenient reference metric says nothing
about solutions far from it. Wang and Ziller~\cite{WZ} and B\"ohm, Wang and
Ziller~\cite{BWZ} characterized the invariant Einstein metrics
variationally, as the critical points of the scalar curvature on invariant
metrics of fixed volume, and compactness results for the unit-volume
invariant Einstein metrics~\cite{BWZ,Graev} concern one homogeneous space
at a time. In the family studied here, both the dimension of the space and
the number of invariant coefficients grow with a parameter $n$, so a
classification for large $n$ needs estimates that are uniform in $n$.

The spaces we study come from the tensor decomposition
$\mathbb C^{2^n}=(\mathbb C^2)^{\otimes n}$ of an $n$-qubit system, and
their invariant metrics have a direct meaning in quantum control: local
Hamiltonians cost nothing, and an interaction acting on a set $S$ of qubits
is charged with a weight that depends only on $|S|$. Khaneja, Brockett and
Glaser~\cite{KBG} reduced time-optimal control of coupled spins with fast
local controls to a geometric problem on a quotient by local operations.
Nielsen and coauthors~\cite{NielsenScience,NDGD} related Riemannian control
costs to quantum circuit size, and Brown~\cite{BrownEquivalence} showed how
this relation depends on the penalties assigned to directions that act on
many qubits. We first classify the invariant Einstein metrics on these
quotients and then determine what their distances measure as control
costs.

Put
\[
 G_n=\PU(2^n),\qquad H_n=\PU(2)^n\rtimes S_n,
 \qquad M_n=H_n\backslash G_n,
\]
where $\PU(D)=U(D)/U(1)$, the factors $\PU(2)$ act on the qubits through
tensor products, and $S_n$ permutes the qubits. The group $G_n$ acts
transitively on $M_n$ from the right. Let $D=2^n$, $X=D^2=4^n$ and
\[
 Q(A,B)=-D^{-1}\Tr(AB)\qquad(A,B\in\mathfrak{su}(D)).
\]
A Pauli string is a tensor product of $n$ matrices from
$\{I_2,\sigma_1,\sigma_2,\sigma_3\}$, and its weight is the number of
factors different from $I_2$. The Pauli strings of positive weight,
multiplied by $i$, form a $Q$-orthonormal basis of $\mathfrak{su}(D)$. The
strings of weight one span the Lie algebra of $H_n$, and its
$Q$-orthogonal complement splits into the spans $\mathfrak m_w$ of the
strings of weight $w$, $2\le w\le n$, with
\[
 m_w=\dim\mathfrak m_w=3^w\binom nw,\qquad
 d=\dim M_n=X-1-3n.
\]
By Section~\ref{sec:geometry}, every $G_n$-invariant metric on $M_n$ has
the form
\[
 g_q=\sum_{w=2}^nq_wQ|_{\mathfrak m_w},\qquad q_w>0.
\]
Let $r_w(q)$ be the eigenvalue of the Ricci endomorphism of $g_q$ on
$\mathfrak m_w$. Up to scaling, the Einstein equation is
\begin{equation}\label{intro:equations}
 q_2=1,\qquad q_w>0,\qquad r_2(q)=\cdots=r_n(q).
\end{equation}
Thus $\dim M_n$ grows like $4^n$, while the invariant metrics modulo
scaling form a space of dimension only $n-2$.

Two profiles serve as references:
\[
 \bar q_w^{\,+}=1,\qquad
 \bar q_w^{\,-}=\begin{cases}1,&w\text{ even},\\
 \tau_n,&w\text{ odd},\end{cases}
 \qquad
 \tau_n=\frac{2^n-2(-1)^n}{3\cdot2^n+2(-1)^n}.
\]
The first is the normal metric. For the second, note that the odd-weight
Pauli strings span a subalgebra of $\mathfrak{su}(D)$, isomorphic to
$\mathfrak{so}(D)$ for even $n$ and to $\mathfrak{sp}(D/2)$ for odd $n$,
which is the fixed algebra of an involution~\cite{Bullock}. Scaling $Q$ on
this subalgebra by $\tau_n$ gives an Einstein metric on $\PU(D)$, an
instance of the constructions of Jensen and D'Atri--Ziller
\cite{Jensen,DZ}, and $\bar q^{\,-}$ lists its coefficients on the
horizontal directions. Neither profile is Einstein on $M_n$: passing to the
quotient adds to the Ricci eigenvalue on $\mathfrak m_w$ the term
$4w\tau/\bar q_w^{\,2}$, which grows with $w$ (here $\tau=1$ for
$\bar q^{\,+}$ and $\tau=\tau_n$ for $\bar q^{\,-}$;
see~\eqref{eq:local-deletion}). Our first theorem shows that both profiles
can be corrected to Einstein metrics on $M_n$, and that for large $n$ there
are no other invariant Einstein metrics.

\begin{theorem}\label{thm:classification}
For all sufficiently large $n$, equation~\eqref{intro:equations} has
exactly two solutions $q^+$ and $q^-$, and
\begin{equation}\label{intro:profiles}
 \max_{2\le w\le n}\left|\log\frac{q_w^\pm}{\bar q_w^\pm}\right|
 =O(n4^{-n}).
\end{equation}
Both metrics $g_{q^\pm}$ have positive Einstein constant and are
nondegenerate critical points of the scalar curvature on invariant metrics
of fixed volume, with coindex zero for $g_{q^+}$ and one for $g_{q^-}$. No
positive multiple of $g_{q^+}$ is isometric to $g_{q^-}$.
\end{theorem}

Here the coindex is the number of positive eigenvalues of the Hessian of
the scalar curvature on the space of $G_n$-invariant metrics of fixed
volume; as in~\cite{Lauret,LauretWill}, stability is understood within this
space. Thus $g_{q^+}$ is a strict local maximum of the scalar curvature
among invariant metrics of the same volume, while $g_{q^-}$ has exactly one
direction in which the scalar curvature increases. The threshold for $n$ in
Theorem~\ref{thm:classification} is not explicit, because the last step of
the proof argues by contradiction along sequences of solutions. The other
ingredients come with explicit thresholds: the local solutions exist and
are locally unique for $n\ge2^{16}$ (Theorem~\ref{thm:local}), and the a
priori bounds of Theorem~\ref{thm:apriori} hold for $n\ge16$ and
$n\ge2^{24}$, respectively.

The heart of the proof is an a priori bound for all solutions
of~\eqref{intro:equations}, not only for those near the references. The
quantities
\[
 Q_j(q)=\frac1d\sum_{w=2}^nm_wq_w^j,\qquad
 K(q)=\frac{\max_wq_w}{\min_wq_w},\qquad
 \MM(q)=Q_2(q)Q_{-2}(q)
\]
are invariant under scaling, and Theorem~\ref{thm:apriori} shows that
$\MM(q)$ and $K(q)$ are at most $\exp(O(\sqrt n))$ for every solution. The
proof starts from the Ricci equation at a weight $h$ where $q$ is smallest.
This equation bounds a logarithmic energy $\Delta_h$, which measures how
much $q$ varies along the brackets that produce a fixed Pauli string of
weight $h$. The Jacobi identity transfers this bound to the brackets that
produce a string of weight three, which join the weights $s$ and $s+2$.
The resulting quadratic form is the Dirichlet form of two birth--death
chains, one on the even and one on the odd weights, with spectral gaps at
least $8/(45n)$. As a consequence, $\log q_w$ has variance $O(n)$ on each parity class
(Lemma~\ref{uc:logvariance}). Fed back into the equation at the minimum,
the variance bound controls the reciprocal second moment $Q_{-2}$
(Lemma~\ref{uc:inverse}). The argument closes because the minimum is
attained on a whole weight orbit, which contains at least $n^2$ strings,
while the directions removed in the quotient contribute only a term of
order $n$. A separate argument, which uses the trace of the Ricci tensor
over the even directions and the equation at the largest odd coefficient,
shows that the typical even and odd coefficients differ by at most a factor
$\exp(O(\sqrt n))$ (Proposition~\ref{uc:geomeans}).

Since $\exp(O(\sqrt n))=e^{o(n)}$, Section~\ref{sec:rigidity} only has to
treat sequences of solutions with $\log K=o(n)$. Most Pauli strings have
weight close to $3n/4$. For a string $P_a$ of weight $w$, let $N_w(s,t)$ be
the number of strings $P_b$ of weight $s$ that anticommute with $P_a$ and
for which $P_aP_b$ has weight $t$. In a window of width $o(n)$ around
$3n/4$, $N_w(s,t)$ is within a factor $F$ of $m_sm_t/X$, the value one
would expect if the weights of $P_b$ and $P_aP_b$ were independent for a
uniformly random $b$ (Lemma~\ref{lem:kernel}); this is a uniform local
limit theorem with exponential tilting, in the spirit of
Daniels~\cite{Daniels} and Richter~\cite{Richter}. When the window is wide,
$F$ is not close to one, and we use the comparison only to prove
$K\le2^{278}F^{57}$ (Proposition~\ref{prop:feedback}). The width of the
window needed in this argument, and hence $F$, is controlled by $K$, so the
inequality can be fed back into itself; this shows that $K$ stays bounded
(Theorem~\ref{thm:subexp}). Once $K$ is bounded, Fourier analysis on
$(\F_2^2)^n$ shows that each Ricci eigenvalue is close to a function of its
own coefficient and of the parity of its weight. This forces the
coefficients to be nearly constant on each parity class, with odd-to-even
ratio close to $1$ or $1/3$. Finally, a contraction argument around each
reference profile gives existence and local uniqueness
(Theorem~\ref{thm:local}).

Each invariant metric also defines a control cost. Decompose a traceless
Hermitian Hamiltonian by tensor support, $A=\sum_{S\ne\varnothing}A_S$,
where $A_S$ acts nontrivially exactly on the qubits in $S$, and put
\begin{equation}\label{intro:cost}
 F_q(A)^2=\sum_{|S|\ge2}q_{|S|}\,D^{-1}\Tr(A_S^2).
\end{equation}
Single-qubit terms are free. By Proposition~\ref{qc:control}, the minimum
of $\int_0^1F_q(A(t))\,dt$ over all controls that steer the identity into
the coset $H_nU$ is the quotient distance
$\mathcal C_q(U)=d_{g_q}(H_n,H_nU)$. Since $H_n$ contains the permutations
of the qubits, $\mathcal C_q(U)$ does not change when $U$ is multiplied on
either side by a local unitary or by a qubit permutation. We write
$\mathcal C_0$ for the cost of the normal metric $g_0=g_{\mathbf 1}$.

\begin{theorem}\label{intro:control}
Let $c_0=\pi/\sqrt3$ and normalize Einstein metrics by $q_2=1$. For all
sufficiently large $n$, every invariant Einstein metric on $M_n$ satisfies
\begin{equation}\label{intro:Einstein-diam}
 \diam(M_n,g_q)\le c_0\bigl(1+O(n4^{-n})\bigr).
\end{equation}
Let $U$ be Haar distributed, and let $\mathcal C$ be $\mathcal C_0$ or
$\mathcal C_{q^+}$. Then $\mathcal C(U)\to c_0$ in probability and in every
moment of fixed positive order, the diameter of the corresponding metric
tends to $c_0$, and for every $\varepsilon>0$ there is $c_\varepsilon>0$
such that $\Pp[\mathcal C(U)\le c_0-\varepsilon]\le e^{-c_\varepsilon4^n}$
for all large $n$.
\end{theorem}

By Theorem~\ref{thm:classification}, the bound~\eqref{intro:Einstein-diam}
covers all invariant Einstein metrics on $M_n$. Gate counts behave very
differently. Consider circuits built from arbitrary two-qubit gates on
arbitrary pairs of qubits, with single-qubit gates free and no ancillas. A
covering argument in the spirit of Knill~\cite{Knill} (see
also~\cite{Shende}) shows that, at any fixed accuracy, a Haar-random target
needs of order $4^n/n$ two-qubit gates, except with probability
exponentially small in $4^n$ (Proposition~\ref{qc:approx-hard}).
Consequently, no polynomial in $n$ and in an Einstein distance can bound
the gate count of every target. For the Jensen-type metric $g_{q^-}$,
typical distances are asymptotically at least $\pi/3$ and at most $c_0$
(Theorem~\ref{qc:typical}).

Large costs arise in a different region of the coefficient cone. For
$T\ge1$ let $g_T$ be the metric with $q_2=1$ and $q_w=T$ for $3\le w\le n$.
Since $T$ multiplies a squared norm, moving in a direction of weight at
least three costs $\sqrt T$ times as much length as moving in a two-body
direction of the same normalized norm.

\begin{theorem}\label{intro:penalty}
For each $n\ge4$, the metric $g_T$ with $T\ge1$ has nonnegative Ricci
curvature exactly when $T\le T_R$, where $T_R$ is the unique root greater
than one of an explicit cubic equation and
$T_R\sim(4^n/(48n))^{1/3}$; no $g_T$ with $T\ge1$ is Einstein. Uniformly
for $1\le T\le T_R$, the median and the mean of the distance to a
Haar-random point, as well as the diameter, are $\Theta(\sqrt T)$ as
$n\to\infty$. For fixed $n$, $d_{g_T}$ increases uniformly to the strictly
two-body distance $d_{\mathrm{cc}}$ as $T\to\infty$, and
\[
 0\le d_{\mathrm{cc}}-d_{g_T}
 \le C_nT^{-1/[2(n-1)]}.
\]
\end{theorem}

For instance, $T=(4^n/(192n))^{1/3}$ gives a metric of positive Ricci
curvature whose typical distances are of order $2^{n/3}n^{-1/6}$. The lower
bound on distances uses Bishop volume comparison, as in Brown's work on
complexity geometry~\cite{BrownVolume}; on $M_n$ we compute the exact
interval of nonnegative Ricci curvature and obtain a matching upper bound.
If higher-body controls are forbidden altogether, the admissible directions
form the two-body distribution $\mathcal D_2$, which is bracket generating
of step $n-1$ and has Hausdorff dimension $(3n/4-1)4^n+1$
(Theorem~\ref{qc:flag}). All metrics with $q_2=1$, including both Einstein
metrics, restrict to the same sub-Riemannian distance $d_{\mathrm{cc}}$ on
$\mathcal D_2$. Thus the two Einstein metrics give every target a
cost of at most $c_0(1+o(1))$, whereas the penalty metrics $g_T$, none of
which is Einstein, have typical costs of order $\sqrt T$, which reaches
$(4^n/n)^{1/6}$ within the range of nonnegative Ricci curvature.

Section~\ref{sec:geometry} introduces Pauli labels and derives the formulas
for the Ricci and scalar curvature. Section~\ref{sec:uniform} proves the a
priori bounds, and Section~\ref{sec:rigidity} completes the classification
and the stability analysis. Section~\ref{sec:control} studies control
distances and circuit size, and Section~\ref{sec:penalty} treats the
penalty metrics and the two-body limit.

\section{Invariant metrics and curvature}\label{sec:geometry}
\subsection{Pauli labels and support modules}
Let $\sigma_0=I_2$ and
\[
 \sigma_1=\begin{pmatrix}0&1\\1&0\end{pmatrix},\qquad
 \sigma_2=\begin{pmatrix}0&-i\\i&0\end{pmatrix},\qquad
 \sigma_3=\begin{pmatrix}1&0\\0&-1\end{pmatrix}.
\]
We label Pauli strings by $\mathbb V_n=(\F_2^2)^n$: at each site, $(0,0)$
stands for $\sigma_0$, and $(1,0)$, $(1,1)$, $(0,1)$ stand for $\sigma_1$,
$\sigma_2$, $\sigma_3$. For $a\in\mathbb V_n$ let $P_a$ be $i$ times the
corresponding tensor product. The elements $P_a$ with $a\ne0$ form a
$Q$-orthonormal basis of $\mathfrak{su}(D)$. The weight $\wt(a)$ is the
number of sites at which $a$ is nonzero. For $a=(x_i,z_i)_{i=1}^n$ and
$b=(x_i',z_i')_{i=1}^n$ put
\[
 [a,b]=\sum_{i=1}^n(x_iz_i'+z_ix_i')\in\F_2.
\]
At a single site, the pairing equals one exactly when the two Pauli
matrices anticommute, and two strings anticommute exactly when this happens
at an odd number of sites. Hence $[a,b]=1$ exactly when $P_a$ and $P_b$
anticommute, and
\[
 [P_a,P_b]=0\quad\text{if }[a,b]=0,\qquad
 [P_a,P_b]=\pm2P_{a+b}\quad\text{if }[a,b]=1.
\]
Only squares of these structure constants enter below, so the signs play no
role. The Lie algebra of the identity component $\PU(2)^n$ of $H_n$ is
spanned by the $P_a$ of weight one, and its $Q$-orthogonal complement is
\[
 \mathfrak m=\bigoplus_{a\in\HH}\mathbb RP_a,\qquad
 \HH=\{a\in\mathbb V_n:\wt(a)\ge2\}.
\]
The elements of $\HH$ are called horizontal labels.

For $S\subseteq\{1,\ldots,n\}$ let $W_S$ be the span of the $P_a$ whose
support $\{i:a_i\ne0\}$ equals $S$. As a representation of
$\PU(2)^n\cong\SO(3)^n$, $W_S$ is the tensor product of the
three-dimensional adjoint representations of the factors indexed by $S$
and the trivial representations of the other factors. It is absolutely
irreducible, since its complexification is an outer tensor product of
irreducible representations of the factors, and different supports give
inequivalent representations, since they involve different sets of
factors. By Schur's lemma, every $\operatorname{Ad}(\PU(2)^n)$-invariant
inner product on $\mathfrak m$ is a multiple of $Q$ on each $W_S$ and makes
different $W_S$ orthogonal. The group $S_n$ permutes the supports of each
cardinality transitively, so invariance under $S_n$ forces the multiple to
depend only on $|S|$. The $G_n$-invariant metrics on $M_n$ correspond to
the $\operatorname{Ad}(H_n)$-invariant inner products on $\mathfrak m$, and
we conclude that they are exactly
\[
 g_q=\sum_{w=2}^n q_wQ|_{\mathfrak m_w},\qquad
 \mathfrak m_w=\bigoplus_{\wt(a)=w}\mathbb RP_a,
 \qquad q_w>0.
\]
Let $g_0=g_{\mathbf1}$ be the normal metric and $V_{0,n}=\Vol(g_0)$. Since
$\dim\mathfrak m_w=m_w$,
\begin{equation}\label{eq:volume}
 \Vol(g_q)=V_{0,n}\prod_{w=2}^n q_w^{m_w/2}.
\end{equation}
A function on labels is called radial if it depends only on the weight. We
write $q_a=q_{\wt(a)}$ for the radial extension of a coefficient vector.

Put $\epsilon=(-1)^n$ and $\chi(a)=(-1)^{\wt(a)}$. Checking one site at a
time gives
\begin{equation}\label{eq:parity-identities}
 \chi(a+b)=\chi(a)\chi(b)(-1)^{[a,b]},\qquad
 \sum_{a\in\mathbb V_n}\chi(a)=\epsilon D,\qquad
 \sum_b\chi(b)(-1)^{[a,b]}=\epsilon D\chi(a).
\end{equation}
By the second identity, $(X+\epsilon D)/2$ labels have even weight, so the
numbers of even and odd horizontal labels are
\begin{equation}\label{eq:parity-masses}
 p=\frac{X+\epsilon D}{2}-1,\qquad
 k=\frac{X-\epsilon D}{2}-3n.
\end{equation}
By the first identity, $P_a\mapsto-\chi(a)P_a$ is an involutive
automorphism of $\mathfrak{su}(D)$. Its fixed algebra is spanned by the
$P_a$ of odd weight and is conjugate to $\mathfrak{so}(D)$ for even $n$
and to $\mathfrak{sp}(D/2)$ for odd $n$~\cite{Bullock}. Rescaling $Q$ on
this fixed algebra gives the two-scale metrics on $\PU(D)$ that serve as
references in Section~\ref{sec:local}.

For functions on labels we use two normalizations,
\[
 \|v\|_{\ell^2}^2=\sum_a|v_a|^2,
 \qquad \|v\|_2^2=X^{-1}\sum_a|v_a|^2.
\]
Graph Laplacians and gradients with respect to label coefficients refer to
the counting measure; expectations $\E$ and Fourier transforms on
$\mathbb V_n$ refer to the uniform probability measure. For vectors indexed
by weights we write $\langle u,v\rangle_m=\sum_{w=2}^nm_wu_wv_w$. Unless
stated otherwise, asymptotic bounds are uniform in the weights, and norms
of operators on weight vectors are induced by $\|\cdot\|_\infty$.

\subsection{Ricci equations and scalar curvature}
For a label $a$ of weight $w$, let $N_w(s,t)$ be the number of labels $b$
of weight $s$ with $[a,b]=1$ and $\wt(a+b)=t$. Suppose that the support of
$b$ meets that of $a$ in $j$ sites and that $b$ differs from $a$ at
$\beta$ of them. Then $\wt(a+b)=w+s-2j+\beta$, the two strings anticommute
exactly when $\beta$ is odd, and at each of the $\beta$ sites there are two
possible letters for $b$. Hence
\begin{equation}\label{eq:counts}
 N_w(s,t)=\sum_j\binom wj\binom{n-w}{s-j}3^{s-j}
 \sum_{\substack{1\le\beta\le j,\ \beta\ \mathrm{odd}\\w+s-2j+\beta=t}}
 \binom j\beta2^\beta,
\end{equation}
where binomial coefficients outside their range are zero.

The brackets that produce a multiple of $P_a$ are those of $P_b$ and $P_c$
with $b+c=a$ and $[b,c]=1$. We call such an ordered pair $(b,c)$ an input
pair of $a$, and $(c,b)$ its reverse. Since $[b,a+b]=[b,a]$, the input
pairs of $a$ are the pairs $(b,a+b)$ with $[a,b]=1$, and $N_w(s,t)$ counts
those whose entries have weights $s$ and $t$. Reversing the pairs, and
counting the triples $(a,b,a+b)$ by $a$ and by $b$, gives
\begin{equation}\label{eq:balance}
 N_w(s,t)=N_w(t,s),\qquad
 m_wN_w(s,t)=m_sN_s(w,t),\qquad
 \sum_{s,t\ge2}N_w(s,t)=\frac X2-4w.
\end{equation}
For the last identity, note that a nonzero label anticommutes with exactly
$X/2$ labels and therefore has $X/2$ input pairs. In $2w$ of them the first
entry has weight one, in another $2w$ the second entry has weight one, and
for $w\ge2$ no input pair has two entries of weight one.

The Killing form of $\mathfrak{su}(D)$ is $B=-2XQ$. Since $G_n$ is compact,
the Ricci curvature of $g_q$ is given by~\cite[Chapter~7]{Besse}
\[
 \Ric(Y,Y)=-\tfrac12B(Y,Y)-\tfrac12\sum_j\bigl|[Y,Y_j]_{\mathfrak m}\bigr|^2
 +\tfrac14\sum_{i,j}g_q\bigl([Y_i,Y_j]_{\mathfrak m},Y\bigr)^2,
\]
where $Y\in\mathfrak m$ and $(Y_j)$ is a $g_q$-orthonormal basis of
$\mathfrak m$. Take $Y=P_a/\sqrt{q_w}$ with $\wt(a)=w$ and $Y_j$ the
normalized $P_b$. Every nonzero bracket of two basis vectors has squared
$Q$-norm four, so the three terms equal $X/q_w$,
$-2\sum N_w(s,t)q_t/(q_wq_s)$ and $\sum N_w(s,t)q_w/(q_sq_t)$. Thus
\begin{align}
 r_w&=\frac X{q_w}-\frac2{q_w}\sum_{s,t\ge2}N_w(s,t)\frac{q_t}{q_s}
       +q_w\sum_{s,t\ge2}\frac{N_w(s,t)}{q_sq_t}\label{eq:ricci}\\
 &=\frac{8w}{q_w}+\sum_{s,t\ge2}N_w(s,t)
 \frac{q_w^2-(q_s-q_t)^2}{q_wq_sq_t};\label{eq:energy}
\end{align}
the second form follows from the first by writing $X=8w+2\sum N_w(s,t)$,
by~\eqref{eq:balance}, and symmetrizing in $s$ and $t$. In label form,
\begin{equation}\label{eq:ricci-label}
 r_a=\frac X{q_a}-\frac2{q_a}\sum_b\frac{q_{a+b}}{q_b}
     +q_a\sum_b\frac1{q_bq_{a+b}},
\end{equation}
where $b$ runs over the first entries of the input pairs $(b,a+b)$ of $a$
with both entries horizontal. We call the equation $r_a=\lambda$ the row of
$a$. A positive radial $q$ with $r_2(q)=\cdots=r_n(q)=:\lambda$ is called an
Einstein solution; these are exactly the invariant Einstein metrics.

Let $M=q_h$ be a largest coefficient. For $0<u,v\le M$, say with $v\le u$,
\[
 M^2-(u-v)^2-uv=(M^2-u^2)+v(u-v)\ge0,
\]
so each term of~\eqref{eq:energy} with $w=h$ is at least $N_h(s,t)/M$.
By~\eqref{eq:balance} this gives $r_h\ge(X/2+4h)/M$. In particular, every Einstein
solution has $\lambda>0$.

An ordered pair $(a,b)$ is admissible if $a,b,a+b\in\HH$ and $[a,b]=1$.
Summing~\eqref{eq:ricci-label} over $a\in\HH$ gives the scalar curvature.
The double sums of the second and third terms both become sums over
admissible pairs: in the third, the pair $(b,a+b)$ is admissible and its
sum is $a$. Therefore
\begin{equation}\label{eq:scalar}
 \Scal(g_q)=X\sum_{a\in\HH}q_a^{-1}
 -\sum_{(a,b)\,\mathrm{admissible}}
       \frac{q_{a+b}}{q_aq_b}.
\end{equation}
The right side is defined for arbitrary positive coefficients
$(q_a)_{a\in\HH}$, radial or not. Differentiating it term by term, with
$r_a$ defined by~\eqref{eq:ricci-label} also for nonradial coefficients,
gives
\begin{equation}\label{eq:variation}
 \frac{\partial\Scal}{\partial\log q_a}=-r_a,
 \qquad
 d\Scal=-\sum_{w=2}^nm_wr_w\,d\log q_w,
\end{equation}
the second identity for radial $q$. Sections~\ref{sec:uniform}
and~\ref{sec:rigidity} use the first identity to vary the even coefficients
independently. By~\eqref{eq:volume},
$d\log\Vol=\frac12\sum_wm_w\,d\log q_w$, so the second identity shows that
the critical points of $\Scal$ on invariant metrics of fixed volume are
exactly the Einstein metrics.

\subsection{Trace and moment estimates}
Write $\ell=\lambda/X$, and let
$U_e^{\mathrm{full}}=p^{-1}\sum_eq_e^{-1}$ be the mean of $1/q$ over the
$p$ horizontal labels of even weight; these are all the nonzero labels of
even weight. Let $\mathcal G$ be the set of ordered pairs $(e,f)$ of
nonzero even labels with $[e,f]=1$ and $e+f\in\HH$; by the first identity
in~\eqref{eq:parity-identities}, $e+f$ is then odd.

\begin{lemma}\label{lem:traces}
Every Einstein solution satisfies
\begin{equation}\label{eq:traces}
 0<\ell<Q_{-1},\qquad \ell Q_1>\tfrac12,
 \qquad \ell<U_e^{\mathrm{full}}.
\end{equation}
\end{lemma}
\begin{proof}
The trace of the Ricci endomorphism is $\lambda d=\Scal(g_q)$. The sum over
admissible pairs in~\eqref{eq:scalar} is positive, so
$\lambda d<X\sum_aq_a^{-1}=XdQ_{-1}$. Together with $\lambda>0$, this is
the first inequality.

For the second, multiply~\eqref{eq:energy} by $q_w$ and sum over all
horizontal labels. Each unordered triple $\{a,b,a+b\}$ of horizontal labels
with $[a,b]=1$ occurs six times, once for each choice of the label whose
row is taken and of the order of the other two. If the coefficients of the
triple are $u,v,z$, these six terms add up to
\[
 2\sum_{\mathrm{cyc}}\frac{u^2-(v-z)^2}{vz}
 =6+\frac{2\mathscr S(u,v,z)}{uvz},
\]
where
\[
 \mathscr S(u,v,z)=u^3+v^3+z^3+3uvz
 -u^2(v+z)-v^2(u+z)-z^2(u+v).
\]
This is the polynomial in Schur's inequality, so $\mathscr S\ge0$; indeed,
for $u\ge v\ge z>0$,
\[
 \mathscr S(u,v,z)=(u-v)^2(u+v-z)+z(u-z)(v-z)\ge0.
\]
By~\eqref{eq:balance}, there are $\sum_a(X/2-4\wt(a))/6$ such triples.
Therefore
\[
 \lambda dQ_1=8\sum_a\wt(a)+\sum_{\{a,b,a+b\}}\Bigl(6+\frac{2\mathscr S}{uvz}\Bigr)
 =\frac{Xd}2+4\sum_a\wt(a)
       +2\sum_{\{a,b,a+b\}}\frac{\mathscr S(q_a,q_b,q_{a+b})}{q_aq_bq_{a+b}}
       >\frac{Xd}2.
\]

For the third inequality, sum the rows of the $p$ even horizontal labels.
By~\eqref{eq:parity-identities}, an input pair of an even label consists of
one even and one odd label, so the input pairs of all even labels are the
pairs $(f,e+f)$ and $(e+f,f)$ with $(e,f)\in\mathcal G$, the row being that
of $e$. In the sum of the rows, the terms $2q_e/(q_fq_{e+f})$ coming from
the last part of~\eqref{eq:ricci-label} cancel the terms
$-2q_f/(q_eq_{e+f})$ coming from the middle part, as one sees by exchanging
$e$ and $f$. What remains is
\begin{equation}\label{eq:even-trace}
 p\lambda=X\sum_{e\ \mathrm{even}}q_e^{-1}
 -2\sum_{(e,f)\in\mathcal G}\frac{q_{e+f}}{q_eq_f}<XpU_e^{\mathrm{full}}.
\end{equation}
\end{proof}

\begin{lemma}\label{lem:moment}
Let $q$ be an Einstein solution with $X\ge1024\MM(q)$. Then
\begin{equation}\label{eq:moment-comparison}
 K(q)<32\MM(q)^{3/2}.
\end{equation}
\end{lemma}
\begin{proof}
Both sides are invariant under scaling, so we may assume $Q_2=1$; put
$c=Q_{-2}^{1/2}=\MM^{1/2}$. By Cauchy--Schwarz $c\ge1$, so the hypothesis
gives $X\ge1024$ and hence $n\ge5$. Extend $f=1/q$ and $g=q$ by zero
outside $\HH$, and define, for functions $u,v$ on $\mathbb V_n$,
\[
 T(u,v)(a)=\frac1X\sum_{b:\,[a,b]=1}u_b v_{a+b},\qquad
 A_a=1-2T(f,g)(a),\qquad B_a=T(f,f)(a).
\]
Reversing input pairs shows $T(u,v)=T(v,u)$. Multiplying the row of $a$,
in the form~\eqref{eq:ricci-label}, by $q_a/X$ turns it into the quadratic
equation
\begin{equation}\label{eq:quadratic}
 \ell q_a=A_a+B_aq_a^2.
\end{equation}

By~\eqref{eq:balance}, $a$ has $R_a=X/2-4\wt(a)$ input pairs with both
entries horizontal. Pairing each with its reverse and using
$q_{a+b}/q_b+q_b/q_{a+b}\ge2$ gives $\sum_bq_{a+b}/q_b\ge R_a$, so
$A_a\le1-2R_a/X=8\wt(a)/X$. By Cauchy--Schwarz,
$T(f,g)(a)\le X^{-1}(dQ_{-2})^{1/2}(dQ_2)^{1/2}\le c$ and
$B_a\le X^{-1}dQ_{-2}\le c^2$. Thus
\begin{equation}\label{eq:A-universal}
 -2c\le A_a\le\frac{8\wt(a)}X,\qquad B_a\le c^2.
\end{equation}
For a lower bound on $B_a$, note that $b$ and $a+b$ each run over distinct
horizontal labels, so
$\sum_bq_bq_{a+b}\le\frac12\sum_b(q_b^2+q_{a+b}^2)\le dQ_2=d$. By
Cauchy--Schwarz, $R_a^2\le\sum_b(q_bq_{a+b})^{-1}\sum_bq_bq_{a+b}$, and
since $d<X$ and $4n\le X/16$,
\[
 B_a\ge\frac{R_a^2}{Xd}\ge\Bigl(\frac{X/2-4n}X\Bigr)^2\ge\frac{49}{256}
 >\frac3{16}.
\]
Lemma~\ref{lem:traces} and Cauchy--Schwarz give
$\frac12\le\frac1{2Q_1}<\ell<Q_{-1}\le c$. Inserting these bounds
into~\eqref{eq:quadratic} gives $\frac3{16}q_a^2-cq_a-2c<0$, and since
$c\ge1$ this forces $q_a<8c$.

It remains to bound the coefficients from below. Suppose
$q_a\le1/(4c^2)$. Then $B_aq_a\le1/4<\ell/2$, and~\eqref{eq:quadratic}
gives $\frac\ell2q_a<(\ell-B_aq_a)q_a=A_a\le8n/X$, so $q_a<32n/X$. On the
other hand, $Q_{-2}\ge m_wq_w^{-2}/d$ for each $w$, that is,
$q_w\ge c^{-1}\sqrt{m_w/d}$. The ratio $m_{w+1}/m_w=3(n-w)/(w+1)$ decreases
in $w$, so the minimum of $m_w$ over $2\le w\le n$ is attained at $w=2$ or
$w=n$, and since $3^n\ge9\binom n2$,
\[
 m_w\ge9\binom n2\ge n^2\qquad(2\le w\le n).
\]
Hence, with $w=\wt(a)$,
$q_a\ge c^{-1}\sqrt{m_w/d}>n/(c\sqrt X)\ge32n/X$, the last step being the
hypothesis $X\ge1024c^2$. This contradiction shows that
$q_a>1/(4c^2)$ for every $a$, and therefore $K<8c\cdot4c^2=32\MM^{3/2}$.
\end{proof}

\section{Uniform bounds for all Einstein metrics}\label{sec:uniform}
This section proves the following a priori estimate. It applies to every
Einstein solution, not only to those constructed in
Section~\ref{sec:local}.

\begin{theorem}\label{thm:apriori}
Every Einstein solution satisfies
\begin{align}
 \MM(q)&<2^{52}\exp(3072\sqrt n),&&n\ge16,\label{intro:moment}\\
 K(q)&<2^{83}\exp(4608\sqrt n),&&n\ge2^{24}.\label{intro:ratio}
\end{align}
\end{theorem}

\subsection{The minimum row and logarithmic variance}\label{sec:log}
Throughout this section $q$ is an Einstein solution. Let $h$ be a weight
with $q_h=\min_wq_w$, and put
\[
 \Delta_h=\sum_{s,t\ge2}N_h(s,t)\frac{(q_s-q_t)^2}{q_sq_t}.
\]
This is the energy term in the row of a label of weight $h$. Since
$q_h^2\le q_sq_t$ and $\lambda>0$, the row~\eqref{eq:energy} with $w=h$
gives
\begin{equation}\label{uc:min}
 \Delta_h=8h+q_h^2\sum_{s,t\ge2}\frac{N_h(s,t)}{q_sq_t}-\lambda q_h
 <8h+\frac X2-4h\le\frac{9X}{16},
\end{equation}
where the last step uses $4h\le4n\le X/16$, valid for $n\ge4$.

Put $x_w=\log q_w$. Let $\mu_e,V_e$ and $\mu_o,V_o$ be the mean and the
variance of $x_w$ over the even and over the odd horizontal weights,
computed with the weights $m_w$ (so that the two classes have total
weights $p$ and $k$), and let
\[
 V=\frac pd V_e+\frac kd V_o
\]
be the average variance within the parity classes.

\begin{lemma}\label{uc:logvariance}
For every Einstein solution with $n\ge4$,
\begin{equation}\label{uc:varDelta}
 V\le\frac{45n}{d}\Delta_h<\frac{48n}{X}\Delta_h,
 \qquad V_e,V_o<68n.
\end{equation}
\end{lemma}
\begin{proof}
\emph{A spectral gap on the weights.} Fix a label of weight three and let
$C_s=N_3(s,s+2)$ be the number of its input pairs whose entries have
weights $s$ and $s+2$. The first entry of such a pair meets the support of
the weight-three label in exactly one site, where the letters differ, so
\eqref{eq:counts} gives
\[
 C_s=6\cdot3^{s-1}\binom{n-3}{s-1},\qquad 2\le s\le n-2.
\]
On the even weights and on the odd weights of $\{2,\ldots,n\}$ consider
the birth--death operators
\[
 (Lv)_s=u_s(v_s-v_{s+2})+d_s(v_s-v_{s-2}),
\]
with rates
\[
 u_s=\frac{C_s}{m_s}=\frac{2s(n-s)(n-s-1)}{n(n-1)(n-2)},\qquad
 d_s=\frac{C_{s-2}}{m_s}=\frac{2s(s-1)(s-2)}{9n(n-1)(n-2)},
\]
where transitions to weights outside $\{2,\ldots,n\}$ are omitted. (This
affects only the bottom odd weight, where the transition from $3$ to $1$ is
dropped; the formulas give $d_2=0$ and $u_{n-1}=u_n=0$ automatically.)
Since $m_su_s=C_s=m_{s+2}d_{s+2}$, the operator $L$ is self-adjoint for the
weights $m_s$, and
\[
 \langle Lv,v\rangle_m=\sum_sC_s(v_{s+2}-v_s)^2.
\]
On a parity class with at least two weights, the chain is a connected
path, so the kernel of $L$ consists of the constants. We show that every
positive eigenvalue $\gamma$ of $L$ is at least $8/(45n)$; the argument is
a finite-path version of the weighted-gradient method discussed
in~\cite{ChafaiJoulin}. Let $Lv=\gamma v$ with $v$ nonconstant, put
$a_s=1/[s(s+1)]$ and $z_s=(v_{s+2}-v_s)/a_s$, and choose an edge
$(s,s+2)$ at which $|z_s|$ is maximal; replacing $v$ by $-v$ if necessary,
$z_s>0$. Subtracting the eigenvalue equation at $s$ from that at $s+2$
gives
\[
 \gamma a_sz_s=(u_s+d_{s+2})a_sz_s-u_{s+2}a_{s+2}z_{s+2}-d_sa_{s-2}z_{s-2},
\]
and since $z_{s\pm2}\le z_s$,
\[
 \gamma\ge u_s+d_{s+2}
 -u_{s+2}\frac{a_{s+2}}{a_s}
 -d_s\frac{a_{s-2}}{a_s}.
\]
For an edge in the interior of the path, the right side equals
\[
 \frac{4s(9n^2+9n-8s^2-32s-24)}
 {9n(n-1)(n-2)(s+3)}\ge\frac8{45n};
\]
indeed, the expression in parentheses decreases in $s$ and equals
$n^2+9n+8$ at $s=n-2$, and $s/(s+3)\ge2/5$. At the lowest edge of each
path the term $d_sa_{s-2}z_{s-2}$ is absent, because there is no edge
below. In the closed formula, this term contributes
$-2s^2(s+1)/[9n(n-1)(n-2)]\le0$ (for $s=2$ after cancelling the factor
$s-2$), so the formula still gives a lower bound. At the top edges the
formula for $u_{s+2}$ already vanishes. This proves the bound on $\gamma$.
Expanding in eigenvectors on each parity path gives
\begin{equation}\label{uc:pathgap}
 \sum_s C_s(v_{s+2}-v_s)^2\ge\frac8{45n}
 \left[\sum_{s\ \mathrm{even}}m_s(v_s-\bar v_e)^2+
 \sum_{s\ \mathrm{odd}}m_s(v_s-\bar v_o)^2\right],
\end{equation}
where $\bar v_e,\bar v_o$ are the $m$-weighted means; a parity class
consisting of a single weight contributes zero to both sides.

\emph{Transfer from the minimum row.} Extend $x$ to weight one by
$x_1=x_h$, and let $Y$ be the linear map of $\mathfrak{su}(D)$ with
$YP_b=x_{\wt(b)}P_b$ for $b\ne0$. For a label $a$ write $\ad_a=\ad(P_a)$.
This operator maps each $P_b$ to $\pm2P_{a+b}$ or to zero, so its operator
norm is two, and $[\ad_a,Y]P_b=\pm2(x_{\wt(b)}-x_{\wt(a+b)})P_{a+b}$ when
$[a,b]=1$. Hence
\[
 \|[\ad_a,Y]\|_{\rm HS}^2=4E_a,\qquad
 E_a=\sum_{b:\,[a,b]=1}(x_{\wt(b)}-x_{\wt(a+b)})^2,
\]
and $E_a$ depends only on $\wt(a)$. If $\wt(a)=h$, the input pairs with an
entry of weight one contribute nothing to $E_a$, because their other entry
has weight $h$ and $x_1=x_h$; this is why we chose $x_1=x_h$. Since
$(q_s-q_t)^2/(q_sq_t)=e^y+e^{-y}-2\ge y^2$ with $y=x_s-x_t$, we get
$E_a\le\Delta_h$.

Now choose anticommuting labels $a,b$ of weight $h$ with $\wt(a+b)=3$: for
$h=2$, take two labels whose supports share one site and whose letters
differ there; for $h\ge3$, take two labels with the same support that
differ at exactly three sites. The Jacobi identity gives
$[\ad_a,\ad_b]=\ad[P_a,P_b]=\pm2\ad_{a+b}$ and
\[
 [\ad_{a+b},Y]=\pm\tfrac12\bigl([\ad_a,[\ad_b,Y]]-[\ad_b,[\ad_a,Y]]\bigr).
\]
Since $\|[\ad_a,Z]\|_{\rm HS}\le2\|\ad_a\|\,\|Z\|_{\rm HS}=4\|Z\|_{\rm HS}$
and $E_a=E_b$, it follows that
$\|[\ad_{a+b},Y]\|_{\rm HS}\le4\|[\ad_a,Y]\|_{\rm HS}$, that is,
$E_{a+b}\le16E_a\le16\Delta_h$. In $E_{a+b}$, the input pairs whose
entries have weights $s$ and $s+2$, in either order, number $2C_s$, and all
other terms are nonnegative. Therefore
\[
 \sum_sC_s(x_{s+2}-x_s)^2\le\tfrac12E_{a+b}\le8\Delta_h.
\]
Applying~\eqref{uc:pathgap} to $v=x$ gives $\frac8{45n}dV\le8\Delta_h$,
that is, $dV\le45n\Delta_h$. Since $45X/d<48$ for $n\ge4$, this is the
first assertion. For the second, $p,k\ge3X/8$ for $n\ge4$,
so~\eqref{uc:min} gives
\[
 V_e,V_o\le\frac{45n\Delta_h}{\min(p,k)}
 <45n\cdot\frac{9X/16}{3X/8}=\frac{135}{2}n<68n.
\]
\end{proof}

\begin{lemma}\label{uc:inverse}
Normalize $\min q=1$, let $E=e^{\mu_e}$ and $O=e^{\mu_o}$ be the
geometric means of the coefficients of the two parity classes, and put
$T=\min(E,O)$. Every Einstein solution with $n\ge16$ satisfies
\begin{equation}\label{uc:inversebound}
 Q_{-2}<4T^{-2}\exp(2\sqrt{96n}).
\end{equation}
\end{lemma}
\begin{proof}
Write $R=Q_{-2}$ and $\mu=\min(\mu_e,\mu_o)=\log T$; note that all
$x_w\ge0$. Let $a_*$ be a label of weight $h$, so $q_{a_*}=1$. The middle
term in~\eqref{uc:min} equals $\sum_b(q_bq_{a_*+b})^{-1}$, summed over the
input pairs $(b,a_*+b)$ with horizontal entries. Since
$(q_bq_{a_*+b})^{-1}\le\frac12(q_b^{-2}+q_{a_*+b}^{-2})$ and reversal maps
these pairs to themselves, this sum is at most
$\sum_{b\in\HH}q_b^{-2}=dR$. Thus $\Delta_h<8n+dR$, and
Lemma~\ref{uc:logvariance} gives
\begin{equation}\label{uc:feedbackvariance}
 V<\frac{48n}X(8n+dR)<\frac{384n^2}{X}+48nR.
\end{equation}
Put $\alpha=\sqrt{96n}$, and regard $x$ as a random variable on $\HH$ with
the uniform distribution, so that $R=\E e^{-2x}$. If $x\le\mu-\alpha$, then
$x$ lies at least $\alpha$ below the mean of its own parity class, so
Chebyshev's inequality gives $\Pp(x\le\mu-\alpha)\le V/\alpha^2$. On this
event $e^{-2x}\le1$, and off it $e^{-2x}<e^{-2\mu+2\alpha}$. Consequently
\[
 R\le\frac{V}{\alpha^2}+e^{-2\mu+2\alpha}
 <\frac{4n}{X}+\frac R2+e^{-2\mu+2\alpha},
 \qquad\text{so}\qquad
 R<\frac{8n}{X}+2e^{-2\mu+2\alpha}.
\]
The minimum is attained on the whole weight orbit of $h$, which has
$m_h\ge n^2$ elements, so $R\ge m_h/d>n^2/X\ge16n/X$ for $n\ge16$. Thus
$8n/X<R/2$, and absorbing this term gives
$R<4e^{-2\mu+2\alpha}=4T^{-2}e^{2\alpha}$.
\end{proof}

\subsection{Comparison of the parity scales}\label{sec:unconditional}
Keep the normalization $\min q=1$ and the notation $E,O,T$ of
Lemma~\ref{uc:inverse}, and put
\begin{equation}\label{uc:typicalscales}
 S=\max(E,O),\qquad L=e^{128\sqrt n}.
\end{equation}
Since $2\sqrt{96n}<256\sqrt n$, Lemma~\ref{uc:inverse}, Cauchy--Schwarz and
Lemma~\ref{lem:traces} give
\begin{equation}\label{uc:firstinverse}
 Q_{-2}<\frac{4L^2}{T^2},\qquad Q_{-1}<\frac{2L}T,\qquad
 \ell<Q_{-1}<\frac{2L}T.
\end{equation}
Call a horizontal label good if $|\log q_a-\mu_\sigma|\le128\sqrt n$,
where $\sigma$ is the parity of $\wt(a)$, and bad otherwise. By
Lemma~\ref{uc:logvariance} and Chebyshev's inequality, the bad labels form
a fraction less than $68/128^2<1/128$ of each parity class, so there are
fewer than $X/128$ of them. Good even coefficients lie in $[E/L,EL]$, and
good odd coefficients lie in $[O/L,OL]$.

\begin{lemma}\label{uc:goodpairs}
Let $n\ge16$. Every even horizontal label $a$ has at least $X/8$ input
pairs $(b,a+b)$ in which $b$ is even and good and $a+b$ is odd and good.
Every odd horizontal label has at least $X/8$ input pairs consisting of two
good even labels and at least $X/8$ input pairs consisting of two good odd
labels.
\end{lemma}
\begin{proof}
First count all input pairs, including those with an entry of weight one.
The number of labels $b$ with $[a,b]=1$ and $\chi(b)=\pm1$ is
$\frac14\sum_b(1-(-1)^{[a,b]})(1\pm\chi(b))$, which~\eqref{eq:parity-identities}
evaluates. An even label has $X/4$ input pairs with even first entry and
$X/4$ with odd first entry. An odd label has $c_+$ input pairs with two
even entries and $c_-$ with two odd entries, where
\[
 c_\pm=\frac{X\pm2\epsilon D}{4}.
\]
For an even label of weight $w$, removing the pairs with an entry of
weight one leaves $X/4-2w$ pairs of the required kind. Removing those with
a bad entry costs less than $X/128$, because a pair is determined by its
even entry and also by its odd entry. For an odd label of weight $w$,
entries of weight one occur only in odd--odd pairs, which leaves $c_+$
even--even and $c_--4w$ odd--odd pairs; here removing the pairs with a bad
entry costs less than $2X/128$ in each case. For $n\ge16$ we have
$c_\pm\ge7X/32$ and $4n<X/64$, so all three counts exceed $X/8$.
\end{proof}

\begin{lemma}\label{uc:maxglobal}
Every Einstein solution with $n\ge16$ satisfies
\begin{equation}\label{uc:maxglobalbound}
 M:=\max_w q_w<64SL^3.
\end{equation}
\end{lemma}
\begin{proof}
We use the coefficients $A_a,B_a$ of~\eqref{eq:quadratic}. For an even
label, the $X/8$ pairs of Lemma~\ref{uc:goodpairs} have
$q_bq_{a+b}\le EL\cdot OL$, so
\[
 B_a\ge\frac1{8EOL^2}.
\]
For an odd label, the good pairs from the parity class whose geometric mean
is $T$ give $B_a\ge1/(8T^2L^2)\ge1/(8EOL^2)$. Now let $a$ be a label with
$q_a=M$. Since $q_{a+b}\le M$,
\[
 A_a=1-\frac2X\sum_b\frac{q_{a+b}}{q_b}
 \ge1-2MQ_{-1},
\]
and~\eqref{eq:quadratic} together with $\ell<Q_{-1}$ gives
$B_aM^2=\ell M-A_a<3MQ_{-1}$. By~\eqref{uc:firstinverse},
\[
 M<24EOL^2Q_{-1}<48\frac{EO}{T}L^3=48SL^3.
\]
\end{proof}

If all odd coefficients were much smaller than the even ones, the even
coefficients would be forced to be nearly equal, and the row of the
largest odd coefficient could not balance. The next lemma makes this
quantitative.

\begin{lemma}\label{uc:paritybridge}
Let $\underline q_e$ and $M_e$ be the smallest and the largest even
coefficient, and $M_o$ the largest odd coefficient. Every Einstein solution
with $n\ge4$ satisfies
\begin{equation}\label{uc:paritybridgebound}
 M_o>\frac{\underline q_e^{\,2}}{16M_e}.
\end{equation}
\end{lemma}
\begin{proof}
By scaling we may assume $\underline q_e=1$; write $A=M_e$ and $M=M_o$.
Regard $\mathcal G$ (defined before Lemma~\ref{lem:traces}) as a graph on
the nonzero even labels, with an edge for every pair in $\mathcal G$.

\emph{Spectral gap.} We claim that the Laplacian $L_{\mathcal G}$
satisfies
$\langle L_{\mathcal G}v,v\rangle\ge\frac X8\|v\|_{\ell^2}^2$ whenever
$\sum_ev_e=0$. Let $\mathscr A$ be the anticommutation matrix on the
nonzero labels, $\mathscr A_{ab}=[a,b]$. Each nonzero label anticommutes
with $X/2$ labels, and two distinct nonzero labels anticommute
simultaneously with exactly $X/4$ labels, so
\[
 \mathscr A^2=(X/4)(I+\mathbf1\mathbf1^{\mathsf T}).
\]
Since $\mathscr A\mathbf1=(X/2)\mathbf1$, the eigenvalues of $\mathscr A$
on $\mathbf1^\perp$ are $\pm D/2$. A vector supported on the even labels
with zero sum lies in $\mathbf1^\perp$, and on the even labels
$\mathscr A$ is the adjacency matrix of a graph of degree $X/4$. It follows that the
Laplacian of that graph is at least $X/4-D/2$ on such vectors. The graph
$\mathcal G$ is obtained by deleting the edges $(e,f)$ with $\wt(e+f)=1$,
at most $2n$ at each vertex, and the Laplacian of the deleted graph has
norm at most $4n$. Thus the gap of $\mathcal G$ is at least
$X/4-D/2-4n\ge X/8$ for $n\ge4$.

\emph{A variational estimate.} Fix the odd coefficients and regard the
scalar curvature~\eqref{eq:scalar} as a function of the even
log-coefficients $y_e=\log q_e$, varied independently. The terms that
depend on them are
\[
 \mathscr V=X\sum_e q_e^{-1},\qquad
 \mathscr U=-2\sum_{(e,f)\in\mathcal G}\frac{q_e}{q_fq_{e+f}},\qquad
 -\mathscr W=-\sum_{(e,f)\in\mathcal G}\frac{q_{e+f}}{q_eq_f}:
\]
the source term, the admissible pairs with one even and one odd entry (and
even sum), and the admissible pairs with two even entries. For any
direction $v$,
\[
 -D^2\mathscr U[v,v]=2\sum_{(e,f)\in\mathcal G}(v_e-v_f)^2\frac{q_e}{q_fq_{e+f}}
 =2\sum_{\{e,f\}}(v_e-v_f)^2\,\frac{q_e/q_f+q_f/q_e}{q_{e+f}}
 \ge\frac4M\langle L_{\mathcal G}v,v\rangle,
\]
where the middle sum runs over unordered edges. By the spectral gap,
$-D^2\mathscr U[v,v]\ge\frac X{2M}\|v\|_{\ell^2}^2$ for every $v$ with zero
sum, at every point. At equal even coefficients the gradient of
$\mathscr U$ vanishes, because each edge occurs in both orientations.

Let $\bar y$ be the mean of the $y_e$, put $x=y-\bar y\mathbf1$, and let $P$
remove the mean. Integrating $D^2\mathscr U$ along the segment from
$\bar y\mathbf1$ to $y$ gives
$\langle P\nabla\mathscr U(y),x\rangle\le-\frac X{2M}\|x\|_{\ell^2}^2$, so
$\|P\nabla\mathscr U(y)\|_{\ell^2}\ge\frac X{2M}\|x\|_{\ell^2}$. At an
Einstein solution,~\eqref{eq:variation} gives $\partial\Scal/\partial y_e=
-\lambda$ for every even $e$, so
$P\nabla(\mathscr V+\mathscr U-\mathscr W)=0$. Since all $q_e\ge1$,
\[
 \Bigl|\frac{\partial\mathscr V}{\partial y_e}\Bigr|=\frac X{q_e}\le X,
 \qquad
 \Bigl|\frac{\partial\mathscr W}{\partial y_e}\Bigr|
 =2\sum_f\frac{q_{e+f}}{q_eq_f}\le\frac{MX}2,
\]
and therefore
\begin{equation}\label{uc:bridgevar}
 \|x\|_{\ell^2}\le\frac{2M}X\Bigl(X+\frac{MX}2\Bigr)\sqrt p
 =2M\Bigl(1+\frac M2\Bigr)\sqrt p.
\end{equation}
Moreover $0<\lambda<XU_e^{\mathrm{full}}\le X$ by~\eqref{eq:even-trace}.

\emph{The row of the largest odd coefficient.} Let $o$ be an odd label with
$q_o=M$. Splitting the input pairs in~\eqref{eq:energy} by parity,
\[
 r_o=r_o^{\rm odd}-\frac{\Delta_o^e}{M}+MZ_o^e,
\]
where
\[
 \Delta_o^e=\sum_{\substack{e\text{ even}\\ [o,e]=1}}\frac{(q_e-q_{o+e})^2}{q_eq_{o+e}},
 \qquad
 Z_o^e=\sum_{\substack{e\text{ even}\\ [o,e]=1}}\frac{1}{q_eq_{o+e}}>0,
\]
and $r_o^{\rm odd}$ consists of the term $8\wt(o)/M$ and the contributions
of the odd--odd input pairs. There are $c_--4\wt(o)$ such pairs (proof of
Lemma~\ref{uc:goodpairs}). Both entries of each have coefficient at most
$M$, so the inequality used to prove $\lambda>0$ gives
\[
 Mr_o^{\rm odd}\ge8\wt(o)+c_--4\wt(o)>c_-\ge\frac{7X}{32}.
\]

Now suppose $M\le1/(16A)$. Then $M\le1/16$, and~\eqref{uc:bridgevar} gives
$\|x\|_{\ell^2}\le3M\sqrt p$. The even log-coefficients lie in
$[0,\log A]$, where the exponential function is $A$-Lipschitz, so
$|q_e-e^{\bar y}|\le A|x_e|$. Using
$(q_e-q_{o+e})^2\le2(q_e-e^{\bar y})^2+2(q_{o+e}-e^{\bar y})^2$,
$q_eq_{o+e}\ge1$, and the fact that $e\mapsto o+e$ permutes the even labels
that anticommute with $o$, we get
\[
 \Delta_o^e\le4A^2\|x\|_{\ell^2}^2\le36A^2M^2p.
\]
Since $\lambda=r_o$ and $Z_o^e>0$, it follows that
$\lambda M>Mr_o^{\rm odd}-\Delta_o^e>\frac{7X}{32}-36A^2M^2p$. With
$\lambda<X$ and $p<X$ this yields $M+36A^2M^2>7/32$. But $M\le1/(16A)$ and
$A\ge1$ make the left side at most $\frac1{16}+\frac{36}{256}=\frac{13}{64}
<\frac7{32}$. Hence $M>1/(16A)$, which is~\eqref{uc:paritybridgebound}
after undoing the scaling.
\end{proof}

\begin{proposition}\label{uc:geomeans}
Every Einstein solution with $n\ge16$ satisfies
\begin{equation}\label{uc:geomeanbound}
 \frac{\max(E,O)}{\min(E,O)}<2^{19}L^8.
\end{equation}
\end{proposition}
\begin{proof}
Let $a$ be an even label with $q_a=\underline q_e$. Its input pairs have
one even and one odd entry. Grouping each pair $(b,a+b)$ with $b$ even
together with its reverse in~\eqref{eq:ricci-label}, the row of $a$ reads
\begin{equation}\label{uc:mineven}
 0<\lambda\underline q_e
 =X-2\sum_{\substack{b\ \mathrm{even},\ [a,b]=1\\ a+b\in\HH}}
 \left(\frac{q_{a+b}}{q_b}+\frac{q_b^2-\underline q_e^{\,2}}{q_bq_{a+b}}\right),
\end{equation}
and both terms in the parentheses are nonnegative.

On the $X/8$ good pairs of Lemma~\ref{uc:goodpairs} we have
$q_{a+b}/q_b\ge O/(EL^2)$. Keeping only these pairs and the first term
gives $X>\frac X4\cdot\frac{O}{EL^2}$, that is, $O/E<4L^2$. If
$\underline q_e\le E/(2L)$, then $q_b\ge E/L\ge2\underline q_e$ on good
pairs, so
\[
 \frac{q_b^2-\underline q_e^{\,2}}{q_bq_{a+b}}\ge\frac34\,\frac{q_b}{q_{a+b}}
 \ge\frac34\,\frac{E}{OL^2},
\]
and keeping the second term gives $E/O<\frac{16}3L^2$. Consequently,
$E/O\ge8L^2$ implies $\underline q_e>E/(2L)$.

Since $O/E<4L^2$, the claimed bound can fail only if $E/O\ge8L^2$, and we
assume this from now on. Then $S=E$, $T=O$, and
\begin{equation}\label{uc:evenfloor}
 \underline q_e>\frac E{2L},\qquad
 U_e^{\rm full}\le\frac1{\underline q_e}<\frac{2L}E,\qquad
 \frac{M_e}{\underline q_e}<\frac{64EL^3}{E/(2L)}=128L^4,
\end{equation}
where the last bound uses Lemma~\ref{uc:maxglobal}. Let $a$ now be an odd
label with $q_a=M_o$, and put $y=M_o/O$. The good odd--odd pairs of
Lemma~\ref{uc:goodpairs} give $B_a\ge1/(8O^2L^2)$. In the sum defining
$A_a$, the even--even input pairs contribute at most
\[
 \frac1X\sum_{b\ \mathrm{even}}\frac{q_{a+b}}{q_b}
 \le\frac pXM_eU_e^{\rm full}\le\frac{M_e}{\underline q_e}<128L^4,
\]
and the odd--odd pairs at most
\[
 \frac1X\sum_{b\ \mathrm{odd}}\frac{q_{a+b}}{q_b}
 \le\frac dXM_oQ_{-1}<\frac{2LM_o}{O}=2Ly,
\]
by~\eqref{uc:firstinverse}. Hence $A_a>1-256L^4-4Ly$. Moreover
$\ell<U_e^{\rm full}<2L/E\le2L/O$ by Lemma~\ref{lem:traces}, so
$\ell M_o<2Ly$. Inserting these bounds into~\eqref{eq:quadratic} gives
\[
 \frac{y^2}{8L^2}\le B_aM_o^2=\ell M_o-A_a<256L^4+6Ly,
 \qquad\text{that is,}\qquad y^2<2048L^6+48L^3y.
\]
If $y\ge128L^3$, then $y^2\ge128L^3y\ge48L^3y+80\cdot128L^6$, which is
impossible. Therefore
\begin{equation}\label{uc:oddmax}
 M_o<128OL^3.
\end{equation}
On the other hand, Lemma~\ref{uc:paritybridge} and~\eqref{uc:evenfloor}
give
\[
 M_o>\frac{\underline q_e}{16M_e/\underline q_e}
 >\frac{E/(2L)}{16\cdot128L^4}=\frac{E}{4096L^5}.
\]
Comparing with~\eqref{uc:oddmax} yields $E/O<2^{19}L^8$, which completes
the proof.
\end{proof}

\begin{proof}[Proof of Theorem~\ref{thm:apriori}]
Normalize $\min q=1$, so that $Q_2\le(\max q)^2$. Lemmas~\ref{uc:inverse}
and~\ref{uc:maxglobal} and Proposition~\ref{uc:geomeans} give
\[
 \MM=Q_2Q_{-2}<(64SL^3)^2\,\frac{4L^2}{T^2}
 =2^{14}L^8\Bigl(\frac ST\Bigr)^2<2^{52}L^{24}.
\]
Since $L^{24}=e^{3072\sqrt n}$, this is~\eqref{intro:moment}.
For~\eqref{intro:ratio} we apply Lemma~\ref{lem:moment}, which requires
$X\ge1024\MM$. By~\eqref{intro:moment} it suffices that
\[
 n\log4>62\log2+3072\sqrt n.
\]
After division by $\sqrt n$, the left side increases and the right side
decreases with $n$; at $n=2^{24}$ the left side exceeds $5600$ and the
right side is less than $3073$. The inequality therefore holds for all
$n\ge2^{24}$, and Lemma~\ref{lem:moment} gives
\[
 K<32\MM^{3/2}<2^{83}L^{36}=2^{83}e^{4608\sqrt n}.
\]
\end{proof}

\section{Asymptotic classification}\label{sec:rigidity}
\subsection{A uniform estimate for the bracket counts}\label{sec:kernel}
Call a triple $(w,s,t)$ of weights compatible if $w+s+t$ is odd. By the
first identity in~\eqref{eq:parity-identities}, $N_w(s,t)=0$ unless
$(w,s,t)$ is compatible. For compatible triples put
\[
 \mathcal K_n(w,s,t)=\frac{X N_w(s,t)}{m_sm_t}.
\]
If $b$ were a uniformly random label and the weights of $b$ and $a+b$ were
independent, the expected number of labels $b$ with $\wt(b)=s$ and
$\wt(a+b)=t$ would be $m_sm_t/X$; thus $\mathcal K_n$ measures the
deviation from independence. By~\eqref{eq:balance}, $\mathcal K_n$ is
symmetric in its three arguments. The next lemma shows that $\mathcal K_n$
is close to one near the typical weight $3n/4$. It is a uniform version of
the local limit theorem with exponential tilting~\cite{Daniels,Richter}.

\begin{lemma}\label{lem:kernel}
There are absolute constants $C,\varepsilon_0>0$ such that, for all
sufficiently large $n$, all $1\le h\le\varepsilon_0 n$, and all compatible
triples of integers with
\[
 |w-3n/4|,\ |s-3n/4|,\ |t-3n/4|\le h,
\]
one has
\begin{equation}\label{eq:kernel-bound}
 |\log\mathcal K_n(w,s,t)|
 \le C\left(n^{-1/2}+\frac hn+\frac{h^3}{n^2}\right).
\end{equation}
\end{lemma}
\begin{proof}
Let $\alpha,\beta$ be independent uniform elements of $\F_2^2$ and let
$v\in\{0,1\}^3$ record which of $\alpha$, $\beta$, $\alpha+\beta$ are
nonzero. Then
\[
\begin{array}{c|ccccc}
 v&000&110&101&011&111\\ \hline
 16\Pp(v)&1&3&3&3&6
\end{array}
\]
Each coordinate of $v$ is Bernoulli$(3/4)$, and the three coordinates are
pairwise independent. The pairing of $\alpha$ and $\beta$ equals one
exactly when $v=111$. Hence, for labels $a,b$, both $[a,b]$ and the parity
of $\wt(a)+\wt(b)+\wt(a+b)$ equal the parity of the number of sites with
$v=111$, and for compatible triples
\[
 m_wN_w(s,t)=[x^wy^sz^t](1+3xy+3xz+3yz+6xyz)^n.
\]
Let $\mathbf S_n$ be the sum of $n$ independent copies of $v$, with
coordinates $\mathbf S_{n,1},\mathbf S_{n,2},\mathbf S_{n,3}$. Since
$\Pp(\mathbf S_n=(w,s,t))=m_wN_w(s,t)/X^2$ and
$\Pp(\mathbf S_{n,1}=w)=m_w/X$,
\begin{equation}\label{eq:prob-ratio}
 \mathcal K_n(w,s,t)=
 \frac{\Pp(\mathbf S_n=(w,s,t))}
 {\Pp(\mathbf S_{n,1}=w)\Pp(\mathbf S_{n,2}=s)\Pp(\mathbf S_{n,3}=t)}.
\end{equation}

Let $p_0=3/4$, let $\kappa(u)=\log((1+3e^u)/4)$ be the cumulant generating
function of a Bernoulli$(3/4)$ variable, and let
$\Lambda(\theta)=\log\E e^{\theta\cdot v}$ for $\theta\in\mathbb R^3$.
Because the coordinates of $v$ are pairwise independent with variance
$3/16$, the functions $\Lambda$ and $\sum_i\kappa(\theta_i)$ have the same
derivatives up to order two at $0$:
\[
 \Lambda(\theta)=\sum_i \kappa(\theta_i)+O(|\theta|^3),
 \qquad D^2\Lambda(0)=\tfrac3{16}I.
\]
All derivatives of $\Lambda$ and $\kappa$ are bounded on a fixed
neighborhood of zero. Write $(w,s,t)=nb$ with $b_i=p_0+z_i/n$ and
$|z_i|\le h$. After reducing $\varepsilon_0$, the inverse function theorem
gives unique solutions $\theta$ of $\nabla\Lambda(\theta)=b$ and
$\theta_i^0$ of $\kappa'(\theta_i^0)=b_i$ near zero. They satisfy
\[
 |\theta|+|\theta^0|=O(h/n),\qquad
 |\theta-\theta^0|=O(h^2/n^2),
\]
the second because $\nabla\Lambda(\theta^0)=b+O(|\theta^0|^2)$. The
quantities $\theta\cdot b-\Lambda(\theta)$ and
$\sum_i(\theta_i^0b_i-\kappa(\theta_i^0))$ are the maxima over the tilt
parameter of the concave functions $\vartheta\cdot b-\Lambda(\vartheta)$
and $\vartheta\cdot b-\sum_i\kappa(\vartheta_i)$. Evaluating each function
at the maximizer of the other and using
$\Lambda-\sum_i\kappa=O(|\vartheta|^3)$ shows that the two maxima differ by
$O(h^3/n^3)$, that is,
\begin{equation}\label{eq:rate-difference}
 \left|n\left(\Lambda(\theta)-\theta\cdot b
       -\sum_i(\kappa(\theta_i^0)-\theta_i^0b_i)\right)\right|
 \le C\frac{h^3}{n^2}.
\end{equation}

Let $\Pp_\theta$ be the law of $\mathbf S_n$ when each copy of $v$ is
tilted by $e^{\theta\cdot v-\Lambda(\theta)}$. Under $\Pp_\theta$ the mean
of $\mathbf S_n$ is $nb$, and
$\Pp(\mathbf S_n=nb)=e^{n(\Lambda(\theta)-\theta\cdot b)}
\Pp_\theta(\mathbf S_n=nb)$. For tilts of size $O(\varepsilon_0)$, all five
atoms of the tilted law of $v$ have probabilities bounded below, so the
tilted covariance matrices are uniformly positive definite. The
differences of the support points generate $\mathbb Z^3$ (subtracting
$110$, $101$, $011$ from $111$ gives the unit vectors), so the tilted
characteristic function has modulus bounded away from one outside any
neighborhood of $0$ in the torus $(\mathbb R/2\pi\mathbb Z)^3$. Near $0$,
after removal of the linear term, its logarithm is
$-\frac12\xi^{\mathsf T}D^2\Lambda(\theta)\xi+O(|\xi|^3)$ with uniform
constants. Fourier inversion, with the torus split at $|\xi|=n^{-2/5}$,
gives
\[
 \Pp_\theta(\mathbf S_n=nb)=
 \frac{1+O(n^{-1/2})}{(2\pi n)^{3/2}\sqrt{\det D^2\Lambda(\theta)}}.
\]
Indeed, on $|\xi|\le n^{-2/5}$ the cubic term changes the integrand by at
most $Cn|\xi|^3e^{-cn|\xi|^2}$, whose integral is $O(n^{-2})$, that is,
$O(n^{-1/2})$ relative to the main term; the rest of the torus contributes
an exponentially small amount. The same argument in dimension one applies
to each marginal, with $\theta_i^0$ in place of $\theta$, and the factors
$(2\pi n)^{3/2}$ cancel in~\eqref{eq:prob-ratio}. Finally,
$\det D^2\Lambda(\theta)$ and $\prod_i\kappa''(\theta_i^0)$ both equal
$(3/16)^3+O(h/n)$. Taking logarithms in~\eqref{eq:prob-ratio} and
using~\eqref{eq:rate-difference} proves~\eqref{eq:kernel-bound}.
\end{proof}

\subsection{Improving the coefficient bound}\label{sec:feedback}
Throughout this subsection we consider a sequence of Einstein solutions
with $n\to\infty$; limits refer to this sequence, and all constants are
absolute. Let $\CC\subseteq\HH$ be a union of weight orbits whose
complement in $\mathbb V_n$ has probability at most $\eta$. The labels in
$\CC$ are called central; let $\CC_e$ and $\CC_o$ be the central labels of
even and of odd weight, and $\rho_e=|\CC_e|/X$, $\rho_o=|\CC_o|/X$. Assume
\begin{equation}\label{eq:rough-hypotheses}
 \rho_e,\rho_o\longrightarrow\tfrac12,\qquad
 \eta K^6\longrightarrow0,\qquad \log K=o(n),
\end{equation}
and that for all compatible triples of weights occurring in $\CC$,
\begin{equation}\label{eq:factor}
 F^{-1}\le\mathcal K_n(w,s,t)\le F,\qquad
 F\ge2,\quad \log F=o(n).
\end{equation}

\begin{proposition}\label{prop:feedback}
Under these assumptions, for all sufficiently large $n$ in the sequence,
\begin{equation}\label{eq:feedback}
 K\le2^{278}F^{57}.
\end{equation}
\end{proposition}
\begin{proof}
Rescale so that the mean of $1/q$ over $\CC_e$ equals one. Then some
central even coefficient is at most one and some is at least one, so
$K^{-1}\le q_a\le K$ for every horizontal $a$. Let $t$ be the mean of
$1/q$ over $\CC_o$, let $v_e$ and $v_o$ be the means of $q$ over $\CC_e$
and $\CC_o$, and put
\[
 M_e=\max_{\CC_e}q,\qquad M_o=\max_{\CC_o}q,\qquad x=M_e,\qquad y=M_ot.
\]
By Cauchy--Schwarz, $v_e\ge1$ and $tv_o\ge1$; hence $x\ge1$ and $y\ge1$.

\emph{Step 1: comparison of the row coefficients.} We use the coefficients
$A_a,B_a$ of~\eqref{eq:quadratic}. Let $a$ be central, and for a function
$u$ on labels write $\Sigma_\sigma(u)=X^{-1}\sum_{b\in\CC_\sigma}u_b$,
$\sigma\in\{e,o\}$. Let $u,v$ be positive radial functions. Grouping the
input pairs of $a$ with central entries by the weights of the entries and
applying~\eqref{eq:factor}, we see that these pairs contribute to
$X^{-1}\sum_bu_bv_{a+b}$ between $F^{-1}$ and $F$ times
$\Sigma_e(u)\Sigma_o(v)+\Sigma_o(u)\Sigma_e(v)$ if $a$ is even, and times
$\Sigma_e(u)\Sigma_e(v)+\Sigma_o(u)\Sigma_o(v)$ if $a$ is odd. For
$u=v=1/q$ these expressions equal $2\rho_e\rho_ot$ and
$\rho_e^2+\rho_o^2t^2$. There are at most $2\eta X$ input
pairs with a noncentral entry, and each term $q_b^{\pm1}q_{a+b}^{\pm1}$ is
at most $K^2$, so these pairs change such sums by at most
$2\eta K^2\to0$. Consequently, for large $n$,
\begin{equation}\label{eq:B-rough}
 B_a\ge\frac{2t}{5F}\quad(a\in\CC_e),\qquad
 B_a\ge\frac{1+t^2}{5F}\quad(a\in\CC_o).
\end{equation}

\emph{Step 2: the central maxima.} By Lemma~\ref{lem:traces},
$\ell<Q_{-1}\le\frac Xd(\rho_e+\rho_ot+\eta K)<\frac35(1+t)$ for large
$n$. Let $a$ be a central label with $q_a=M:=\max(M_e,M_o)$. Then
\[
 A_a\ge1-2M(\rho_e+\rho_ot)-4\eta K^2
       \ge\tfrac12-\tfrac{11}{10}M(1+t),
\]
and~\eqref{eq:quadratic} gives $B_aM^2=\ell M-A_a<2M(1+t)$.
By~\eqref{eq:B-rough} and $(1+t)^2\le2(1+t^2)$,
\begin{equation}\label{eq:max-cases}
 M=M_e\ \Longrightarrow\ x<5F(1+t^{-1}),\qquad
 M=M_o\ \Longrightarrow\ M_o<\frac{20F}{1+t}.
\end{equation}
At a central odd label with $q_a=M_o$, both entries of each input pair
have the same parity, so $A_a\ge\tfrac12-\tfrac{11}{10}(x+y)$,
and~\eqref{eq:quadratic} gives
\begin{equation}\label{eq:xy}
 \frac{1+t^{-2}}{5F}\,y^2
 <\left(\frac{17}{10}+\frac3{5t}\right)y+\frac{11}{10}x.
\end{equation}
We now bound $x$, $y$ and $t$. If $M=M_o$, then~\eqref{eq:max-cases}
gives $x\le M_o<20F$ and $y<20Ft/(1+t)<20F$; if moreover $t\le1$, then
$1\le y<20Ft$, so $t>1/(20F)$. If $M=M_e$ and $t\ge1$, then $x<10F$,
and~\eqref{eq:xy} gives $y^2/(5F)<\frac{23}{10}y+11F$, which is impossible
for $y\ge20F$. If $M=M_e$ and $t\le1$, then $x<5F(1+t^{-1})$ and
$y=M_ot\le M_et<5F(1+t)\le10F$. In this last case, divide~\eqref{eq:xy} by
$y$, use $y\ge1$ on the left and $x/y\le x<5F(1+t^{-1})$ on the right, and
multiply by $5Ft^2$; since $F\ge2$, this gives
\[
 1+t^2<36F^2t^2+\tfrac{61}{2}F^2t.
\]
The right side increases with $t$ and is less than one at
$t=1/(64F^2)$, so $t>1/(64F^2)$ and $x<5F(1+64F^2)$. In all cases,
\begin{equation}\label{eq:central-bounds}
 x<512F^3,\qquad y<32F,\qquad t>\frac1{128F^2},\qquad
 \max_{\CC}q=\max\Bigl(x,\frac yt\Bigr)<2^{12}F^3.
\end{equation}
The noncentral labels have mass at most $\eta$ and coefficients at most
$K$, so $Q_1<2^{13}F^3$ for large $n$. Also
$U_e^{\rm full}\le(X\rho_e+\eta XK)/p<2$, so Lemma~\ref{lem:traces}
gives
\begin{equation}\label{eq:ell-rough}
 2^{-14}F^{-3}<\ell<2.
\end{equation}

\emph{Step 3: an upper bound for $t$.} Suppose first that $t\ge8F$. For a
central even label $a$, the input pairs with central entries give
\[
 A_a\le1-\frac{2\rho_e\rho_o\,tv_e}{F}\le-\frac{tv_e}{4F},
\]
because $\rho_e\rho_o\ge(49/100)^2$ for large $n$, $v_e\ge1$ and
$tv_e/F\ge8$; moreover $B_a\le Ft$ for large $n$. Since $\ell q_a>0$,
\eqref{eq:quadratic} gives $B_aq_a^2>-A_a$, and therefore
\begin{equation}\label{eq:even-floor}
 q_a>\frac1{2F}\qquad(a\in\CC_e).
\end{equation}

We claim that $t\le2^{30}F^6$ for large $n$. Suppose instead that
$t>2^{30}F^6$ along a subsequence. Then~\eqref{eq:even-floor} holds,
and~\eqref{eq:central-bounds} gives $M_o=y/t<2^{-25}F^{-5}$; in particular
$M_oF\le1$.

Let $\mathcal G_\CC$ be the graph on $\CC_e$ whose ordered edges are the
pairs $(e,f)$ with $[e,f]=1$ and $e+f\in\CC$ (then $e+f\in\CC_o$). For
large $n$ its Laplacian satisfies
$\langle L_{\mathcal G_\CC}v,v\rangle\ge\frac X{16}\|v\|_{\ell^2}^2$ when
$\sum_ev_e=0$. To see this, extend $v$ by zero to all nonzero even labels
and compare with the graph of all pairs of even labels with $[e,f]=1$,
whose Laplacian is at least $X/4-D/2$ on such vectors (proof of
Lemma~\ref{uc:paritybridge}). Each central label has at most $\eta X$
neighbors outside $\CC_e$, and at most $\eta X$ neighbors $f\in\CC_e$
with $e+f\notin\CC$. The edges of the first kind contribute at most
$\eta X\|v\|_{\ell^2}^2$ to the Laplacian form, and those of the second
kind at most $2\eta X\|v\|_{\ell^2}^2$. The gap is therefore at least
$X/4-D/2-3\eta X\ge X/16$.

Fix the odd coefficients and vary the central even log-coefficients; let
$z_e=\log q_e-\overline{\log q}$ for $e\in\CC_e$, where the bar denotes the
mean over $\CC_e$. The part of~\eqref{eq:scalar} that depends on these
coefficients and involves only central labels is
$\mathscr V_\CC+\mathscr U_\CC-\mathscr W_\CC$, where
\[
 \mathscr V_\CC=X\sum_{e\in\CC_e}q_e^{-1},\qquad
 \mathscr U_\CC=-2\sum_{(e,f)\in\mathcal G_\CC}\frac{q_e}{q_fq_{e+f}},\qquad
 \mathscr W_\CC=\sum_{(e,f)\in\mathcal G_\CC}\frac{q_{e+f}}{q_eq_f}.
\]
As in the proof of Lemma~\ref{uc:paritybridge},
$-D^2\mathscr U_\CC[v,v]\ge\frac4{M_o}\langle L_{\mathcal G_\CC}v,v\rangle
\ge\frac X{4M_o}\|v\|_{\ell^2}^2$ for $v$ with zero sum, and the gradient
of $\mathscr U_\CC$ vanishes at equal even coefficients. Integrating along
the segment from equal coefficients to the actual ones gives
\begin{equation}\label{eq:coercive}
 \frac X{4M_o}\|z\|_{\ell^2}\le\|P\nabla\mathscr U_\CC\|_{\ell^2},
\end{equation}
where $P$ subtracts the mean over $\CC_e$. At an Einstein solution, $P$
annihilates the gradient of the full scalar curvature with respect to the
central even log-coefficients, by~\eqref{eq:variation}. This gradient
differs from that of $\mathscr V_\CC+\mathscr U_\CC-\mathscr W_\CC$ by the
contribution of the admissible pairs that involve a noncentral label. A
central label lies in at most $6\eta X$ such pairs, and each corresponding
term is at most $K^3$, so this contribution is at most $6\eta XK^3$ in
each coordinate. By~\eqref{eq:central-bounds} and~\eqref{eq:even-floor},
\[
 \Bigl|\frac{\partial\mathscr V_\CC}{\partial z_e}\Bigr|=\frac X{q_e}<2FX,
 \qquad
 \Bigl|\frac{\partial\mathscr W_\CC}{\partial z_e}\Bigr|
 \le2\cdot\frac X4\cdot M_o(2F)^2=2M_oF^2X.
\]
Since $\eta K^3\to0$ and $M_oF\le1$, \eqref{eq:coercive} yields
\begin{equation}\label{eq:z-bound}
 \frac{\|z\|_{\ell^2}}{\sqrt{|\CC_e|}}
 \le4M_o\bigl(2F+2M_oF^2+6\eta K^3\bigr)
 \le32M_oF
\end{equation}
for large $n$.

Let $o\in\CC_o$ satisfy $q_o=M_o$, and split its row as in the proof of
Lemma~\ref{uc:paritybridge}:
\[
 r_o=r_o^{\mathrm{odd}}-\frac{\Delta_o^e}{M_o}+M_oZ_o^e,\qquad Z_o^e>0.
\]
For input pairs $(e,o+e)$ with both entries central,
$q_eq_{o+e}\ge(2F)^{-2}$ by~\eqref{eq:even-floor}, and the exponential
function is $512F^3$-Lipschitz on the range of the central even
log-coefficients by~\eqref{eq:central-bounds}. Comparing $q_e$ and
$q_{o+e}$ with the geometric mean of the central even coefficients, as in
the proof of Lemma~\ref{uc:paritybridge}, the central pairs contribute to
$\Delta_o^e$ at most
\[
 4(2F)^2(512F^3)^2\|z\|_{\ell^2}^2
 \le2^{32}M_o^2F^{10}X<2^{-18}X.
\]
The remaining input pairs, at most $2\eta X$ of them, contribute at most
$2K^2$ each. Hence, for large $n$,
\begin{equation}\label{eq:even-defect-small}
 \Delta_o^e\le\frac X{32}.
\end{equation}
On the other hand, $o$ has $c_--4\wt(o)$ odd--odd input pairs, where
$c_-=(X-2\epsilon D)/4\ge7X/32$. In $M_or_o^{\mathrm{odd}}$, each of them
with both entries central contributes at least one, because both
coefficients are at most $M_o$, and each of the others contributes at
least $-2K^2$. Hence
\[
 M_or_o^{\mathrm{odd}}
 \ge c_--2\eta X-4\eta XK^2>\frac X5
\]
for large $n$. Together with~\eqref{eq:even-defect-small}, this gives
$\lambda M_o=M_or_o>M_or_o^{\mathrm{odd}}-\Delta_o^e>X/8$, that is,
$\ell M_o>1/8$. This contradicts $\ell<2$ and $M_o<2^{-25}$, and proves
\begin{equation}\label{eq:t-upper}
 t\le2^{30}F^6.
\end{equation}

\emph{Step 4: a lower bound for the central coefficients.}
By~\eqref{eq:t-upper}, the upper comparison in Step~1 gives
\[
 B_a\le F(1+t^2)+2\eta K^2\le2^{62}F^{13}\qquad(a\in\CC).
\]
Put $\ell_0=2^{-14}F^{-3}$ and $q_0=2^{-77}F^{-16}$. If a central label had
$q_a\le q_0$, then $B_aq_a\le\ell_0/2<\ell/2$ by~\eqref{eq:ell-rough},
and~\eqref{eq:quadratic} together with $A_a\le8\wt(a)/X$ would give
\[
 q_a<\frac{2A_a}{\ell}\le\frac{16n}{\ell_0 X}=\frac{2^{18}nF^3}{X}.
\]
This is impossible for large $n$, because $q_a\ge K^{-1}$ and
$nKF^3/X\to0$ by~\eqref{eq:rough-hypotheses} and~\eqref{eq:factor}.
Therefore
\begin{equation}\label{eq:core-box}
 2^{-77}F^{-16}<q_a<2^{12}F^3\qquad(a\in\CC).
\end{equation}

\emph{Step 5: conclusion.} The noncentral labels have mass at most $\eta$
and satisfy $K^{-1}\le q\le K$, so~\eqref{eq:core-box} gives
$Q_2\le2^{26}F^6$ and $Q_{-2}\le2^{156}F^{32}$ for large $n$. Thus
$\MM\le2^{182}F^{38}$. Since $\log F=o(n)$, eventually $X\ge1024\MM$, and
Lemma~\ref{lem:moment} gives
\[
 K<32(2^{182}F^{38})^{3/2}=2^{278}F^{57}.
\]
\end{proof}

\subsection{Uniform convergence of all coefficients}\label{sec:bounded}

\begin{theorem}\label{thm:subexp}
Let $q^{(n)}$ be Einstein solutions, for $n$ running through a sequence
tending to infinity, with $\log K(q^{(n)})=o(n)$. Then $K(q^{(n)})$ is
bounded, and with the normalization $q_2=1$,
\[
 \min_{\tau\in\{1,1/3\}}\max_{2\le w\le n}
 \left|q_w-\begin{cases}1,&w\text{ even}\\\tau,&w\text{ odd}\end{cases}\right|\longrightarrow0.
\]
\end{theorem}
\begin{proof}
\emph{Boundedness of $K$.} Put
\[
 s_n=\log K(q^{(n)}),\qquad \beta_n=s_n+\log n,\qquad
 h_n=8\sqrt{n\beta_n},
\]
and let $\CC$ consist of the labels with $|\wt(a)-3n/4|\le h_n$. Since
$s_n=o(n)$, also $h_n=o(n)$; hence $\CC\subseteq\HH$ for large $n$, and
Lemma~\ref{lem:kernel} applies with $h=h_n$. The weight of a uniformly
random label is binomial with parameters $n$ and $3/4$, so Hoeffding's
inequality~\cite{Hoeffding} gives
\begin{equation}\label{eq:tail}
 \eta_n:=\Pp(|\wt-3n/4|>h_n)
 \le2e^{-2h_n^2/n}=2K^{-128}n^{-128}.
\end{equation}
The parity masses of $\CC$ tend to $1/2$, because $\E\chi=(-1)^n2^{-n}$
by~\eqref{eq:parity-identities} and $\eta_n\to0$. By
Lemma~\ref{lem:kernel},~\eqref{eq:factor} holds with
\begin{equation}\label{eq:F-choice}
 F_n=2\exp\left[C\left(n^{-1/2}+\frac{h_n}{n}+\frac{h_n^3}{n^2}\right)\right].
\end{equation}
Since $h_n/n=8\sqrt{\beta_n/n}$, $h_n^3/n^2=512\beta_n^{3/2}n^{-1/2}$, and
$\beta_n\to\infty$ with $\beta_n=o(n)$,
\begin{equation}\label{eq:logF}
 \log F_n=\log2+O\left(n^{-1/2}+\sqrt{\beta_n/n}+\beta_n^{3/2}n^{-1/2}\right)
 =o(\beta_n).
\end{equation}
In particular $\log F_n=o(n)$, and~\eqref{eq:tail} gives $\eta_nK^6\to0$.
Thus Proposition~\ref{prop:feedback} applies and yields, for large $n$,
\begin{equation}\label{eq:log-feedback}
 s_n\le278\log2+57\log F_n.
\end{equation}
Suppose that $s_n$ is unbounded, and pass to a subsequence with
$s_n\to\infty$. For indices with $s_n\ge\log n$ we have $\beta_n\le2s_n$,
so~\eqref{eq:logF} gives $\log F_n=o(s_n)$. For indices with
$s_n<\log n$ we have $\beta_n<2\log n$, so $h_n^3/n^2\to0$ and
$F_n\to2$. In both cases~\eqref{eq:log-feedback} contradicts
$s_n\to\infty$. Hence $K(q^{(n)})$ is bounded. Then $\beta_n=O(\log n)$
and $F_n\to2$, so~\eqref{eq:log-feedback} even gives
\begin{equation}\label{eq:universal-bound}
 \limsup_{n\to\infty}K(q^{(n)})\le2^{335}.
\end{equation}

\emph{Approximation of the rows by parity-constant functions.} Normalize
now $\max_wq_w=1$, and let $K_*$ be a bound for $K(q^{(n)})$, so that
$K_*^{-1}\le q_w\le1$. Extend $f=1/q$ and $g=q$ by zero outside $\HH$. On
$\mathbb V_n$ with the uniform probability measure put
\[
 \widehat u(\xi)=X^{-1}\sum_a u(a)(-1)^{[\xi,a]},\qquad
 (u*v)(a)=X^{-1}\sum_b u(b)v(a+b).
\]
Then $\widehat{u*v}=\widehat u\,\widehat v$ and
$\sum_\xi|\widehat u(\xi)|^2=\|u\|_2^2$. The symplectic maps of $\F_2^2$
at each site, together with the permutations of the sites, preserve the
pairing and act transitively on each weight orbit, so $\widehat u$ is
radial when $u$ is. Every nonzero weight orbit has $3^j\binom nj\ge3n$
elements, and therefore $|\widehat u(\xi)|\le\|u\|_2/\sqrt{3n}$ for
$\xi\ne0$ and radial $u$. Since $\widehat{u*v}(0)=\E u\,\E v$, Parseval's
identity gives, for radial $u$ and arbitrary $v$,
\begin{equation}\label{eq:Fourier}
 \|u*v-\E u\,\E v\|_2
 \le\frac{\|u\|_2\|v\|_2}{\sqrt{3n}}.
\end{equation}
Since $(-1)^{[a,b]}=\chi(a)\chi(b)\chi(a+b)$
by~\eqref{eq:parity-identities}, the operator $T$ of
Lemma~\ref{lem:moment} can be written as
\[
 T(u,v)=\tfrac12\{u*v-\chi[(\chi u)*(\chi v)]\}.
\]
Let $\rho_e,\rho_o$ be the proportions of even and odd horizontal labels
among all $X$ labels, and let $U_\sigma$ and $v_\sigma$ be the means of
$1/q$ and of $q$ over the horizontal labels of parity $\sigma$. Replacing
each convolution in $T(f,g)$ and $T(f,f)$ by the product of the means
turns $A_a$ and $B_a$ into
\begin{align*}
 A_e^0&=1-2\rho_e\rho_o(U_ev_o+U_ov_e),&
 B_e^0&=2\rho_e\rho_oU_eU_o,\\
 A_o^0&=1-2\rho_e^2U_ev_e-2\rho_o^2U_ov_o,&
 B_o^0&=\rho_e^2U_e^2+\rho_o^2U_o^2,
\end{align*}
for labels of even and odd weight, respectively. All these numbers are
bounded. Since $f$ and $g$ are bounded radial functions,~\eqref{eq:Fourier}
shows that $\|r/X-F_\sigma(q)\|_2\to0$, where
\[
 F_\sigma(z)=\frac{A_\sigma^0}z+B_\sigma^0z
\]
and $\sigma$ is the parity of the label. By Cauchy--Schwarz within each
class, $U_\sigma v_\sigma\ge1$; together with
$\rho_e+\rho_o=d/X=1-O(n/X)$ and $(\rho_e-\rho_o)^2=O(1/X)$, this gives
$A_\sigma^0\le O(n/X)$. Moreover $U_\sigma\ge1$, so $B_\sigma^0\ge1/4$ for
large $n$, and hence $F_\sigma'\ge1/8$ on $[K_*^{-1},1]$. Since
$r_a/X=\ell$ for every $a$, we get $\|\ell-F_\sigma(q)\|_2\to0$. For two
independent uniform labels $a,a'$ of the same parity,
$|q_a-q_{a'}|\le8|F_\sigma(q_a)-F_\sigma(q_{a'})|$; averaging the squares
shows that the variance of $q$ on each parity class tends to zero.

\emph{Uniform convergence.} Let $\tilde q_e,\tilde q_o$ be the means of $q$
on the two parity classes; they lie in $[K_*^{-1},1]$. Let $\tilde g$ equal
$\tilde q_e$ on all even labels and $\tilde q_o$ on all odd labels,
including those of weight zero and one, and put $\tilde f=1/\tilde g$. The
labels of weight at most one have mass $O(n/X)$, so the variance bound
gives
\[
 \|f-\tilde f\|_2+\|g-\tilde g\|_2\longrightarrow0.
\]
For arbitrary functions,
\begin{equation}\label{eq:T-Cauchy}
 \|T(u,v)\|_\infty\le\|u\|_2\|v\|_2,
\end{equation}
by Cauchy--Schwarz. By bilinearity, $T(f,g)-T(\tilde f,\tilde g)$ and
$T(f,f)-T(\tilde f,\tilde f)$ tend to zero uniformly on all labels, so the
row coefficients converge uniformly and not only in $L^2$. The functions
$T(\tilde f,\tilde g)$ and $T(\tilde f,\tilde f)$ depend only on the parity
of the label, and the counts $X/4$ and $c_\pm$ from the proof of
Lemma~\ref{uc:goodpairs} give them explicitly. With
$t=\tilde q_o/\tilde q_e$, the result is
$\sup_a|r_a/X-F_\sigma^*(q_a)|\to0$, where
\[
 F_e^*(z)=\frac{1-(t+t^{-1})/2}{z}
                    +\frac{z}{2\tilde q_e\tilde q_o},
 \qquad
 F_o^*(z)=\left(\frac{1/4-\epsilon/(2D)}{\tilde q_o^{\,2}}
              +\frac{1/4+\epsilon/(2D)}{\tilde q_e^{\,2}}\right)z.
\]
Since $1-(t+t^{-1})/2\le0$, both functions have derivative bounded below by
a positive constant on the coefficient range. As $r_a/X=\ell$ for all $a$,
all coefficients of the same parity are therefore uniformly close to one
another, and hence to $\tilde q_e$ or $\tilde q_o$. In particular
$F_e^*(\tilde q_e)$ and $F_o^*(\tilde q_o)$ both equal $\ell+o(1)$;
multiplying their difference by $-4t\tilde q_e$ gives
\[
 (3+\beta)t^2-4t+1-\beta=o(1),\qquad \beta=\frac{2\epsilon}{D}.
\]
The left side equals $(3+\beta)(t-1)(t-\tau_n)$, and $\tau_n\to1/3$, so
$\operatorname{dist}(t,\{1,1/3\})\to0$. Dividing by $q_2$, which is
uniformly close to $\tilde q_e$, proves the theorem.
\end{proof}

\subsection{Local existence and uniqueness}\label{sec:local}
The fixed-volume logarithmic coordinates and the corresponding projection
are
\[
 E_n=\Bigl\{z\in\mathbb R^{n-1}:\sum_{w=2}^nm_wz_w=0\Bigr\},\qquad
 \Pi_n z=z-\frac1d\Bigl(\sum_wm_wz_w\Bigr)\mathbf1.
\]
In this subsection, norms of vectors, linear maps and bilinear maps are
maximum norms; in particular $\|\Pi_n\|\le2$. For $n\ge4$ the Jensen ratio
satisfies $1/4\le\tau_n\le1/2$ and $|\tau_n-1/3|\le D^{-1}$.

For $z\in E_n$ let $\bar z_e$ and $\bar z_o$ be its $m$-weighted means over
the even and over the odd horizontal weights. Write
\[
 z=W_ez+W_oz+Cz,
\]
where $Cz$ equals $\bar z_e$ on even weights and $\bar z_o$ on odd weights,
and $W_ez$, $W_oz$ are the even and odd parts of $z-Cz$. These are
complementary projections that preserve $E_n$, with norms at most $2$, $2$
and $1$. Put
\begin{equation}\label{eq:model-operators}
 L_1=\tfrac12 I,
 \qquad L_J=\tfrac{13}{6}W_e+\tfrac56W_o-\tfrac12C.
\end{equation}

\begin{lemma}\label{lem:local-linear}
Let $\bar q$ be one of the reference profiles, and let $L=L_1$ if
$\bar q=\bar q^{\,+}$ and $L=L_J$ if $\bar q=\bar q^{\,-}$. Define
\[
 \Phi(z)=\Pi_n r(\bar q e^z)/X,\qquad z\in E_n,
\]
so that the zeros of $\Phi$ are exactly the Einstein solutions of the form
$\bar qe^z$ with $z\in E_n$. Then
\begin{equation}\label{eq:local-estimates}
 \|L^{-1}\|\le6,\qquad
 \|D\Phi(0)-L\|\le512/n,\qquad
 \|\Phi(0)\|\le32n/X,
\end{equation}
and on $\|z\|\le1/10$,
\begin{equation}\label{eq:local-second}
 \|D^2\Phi(z)\|\le1024.
\end{equation}
\end{lemma}
\begin{proof}
\emph{Inverses.} Clearly $\|L_1^{-1}\|=2$, and
\[
 L_J^{-1}=\tfrac6{13}W_e+\tfrac65W_o-2C,
 \qquad \|L_J^{-1}\|\le\tfrac{12}{13}+\tfrac{12}{5}+2<6.
\]

\emph{Two tools.} Extend radial functions on horizontal labels by zero to
the labels of weight at most one. Let $h$ be radial, and let
$k_0=k_{\mathrm{av}}+k_{\mathrm{dif}}\chi$ be a parity-constant function
with values $k_e,k_o$, so that $k_{\mathrm{av}}=(k_e+k_o)/2$ and
$k_{\mathrm{dif}}=(k_e-k_o)/2$. Writing the condition $[a,b]=1$ as
$\frac12(1-(-1)^{[a,b]})$ and using~\eqref{eq:parity-identities}, one finds
\begin{equation}\label{eq:linear-T}
 T(h,k_0)(a)=\tfrac12 k_{\mathrm{av}}(\E h-\widehat h(a))
 -\tfrac12 k_{\mathrm{dif}}\chi(a)(\E(\chi h)-\widehat{\chi h}(a)).
\end{equation}
For a horizontal label $a$, the weight orbit of $a$ has $m_w\ge n^2$
elements, so Parseval's identity and radial symmetry give
\begin{equation}\label{eq:linear-Fourier}
 |\widehat h(a)|,\ |\widehat{\chi h}(a)|\le\|h\|_\infty/n.
\end{equation}
Hence, if $k_0$ is positive, dropping the two Fourier terms
in~\eqref{eq:linear-T} changes $T(h,k_0)(a)$ by at most
$\frac12\|k_0\|_\infty\|h\|_\infty/n$. Next, let $\bar h_e,\bar h_o$ be
the means of $h$ over the even and odd horizontal labels. The proportions
of even and odd horizontal labels among all labels differ from $1/2$ by at
most $\delta_n=1/(2D)+3n/X\le1/n$. Replacing $\E h$ by
$\frac12(\bar h_e+\bar h_o)$ and $\E(\chi h)$ by
$\frac12(\bar h_e-\bar h_o)$ therefore changes these means by at most
$\delta_n(|\bar h_e|+|\bar h_o|)$. After both replacements,
\eqref{eq:linear-T} becomes
\begin{equation}\label{eq:model-T}
 T(h,k_0)\approx\tfrac14(k_e\bar h_o+k_o\bar h_e)\ \text{on even labels},
 \qquad
 T(h,k_0)\approx\tfrac14(k_e\bar h_e+k_o\bar h_o)\ \text{on odd labels}.
\end{equation}

\emph{Linearization on the full group.} Let $r^G$ be the Ricci row
expression of the group $G_n$, in which input pairs may contain labels of
weight one. Give every label of weight one the reference coefficient
$\tau$, where $\tau=1$ for $\bar q^{\,+}$ and $\tau=\tau_n$ for
$\bar q^{\,-}$, and do not vary it. Put $\bar f=1/\bar q$ and
$\bar g=\bar q$ on all nonzero labels, so that $\bar f$ takes the values
$1,\tau^{-1}$ and $\bar g$ the values $1,\tau$ on even and odd labels. In
the notation of Lemma~\ref{lem:moment}, $r^G/X=f-2fT(f,g)+gT(f,f)$ with
$f=1/q$ and $g=q$. Differentiating at $z=0$ and using $T(u,v)=T(v,u)$
gives
\begin{align}\label{eq:full-linear}
 D(r^G/X)z={}&[-\bar f+2\bar fT(\bar f,\bar g)+\bar gT(\bar f,\bar f)]z\\
 &+2\bar fT(\bar fz,\bar g)-2\bar fT(\bar gz,\bar f)-2\bar gT(\bar fz,\bar f).\notag
\end{align}
Here $\|\bar f\|_\infty\le4$ and $\|\bar g\|_\infty\le1$.

In the first line all arguments of $T$ are parity constant, and the counts
$X/4$ and $c_\pm$ from the proof of Lemma~\ref{uc:goodpairs} give the values
exactly. They agree with~\eqref{eq:model-T}, except that in
$T(\bar f,\bar f)$ on odd labels the weights $c_\pm/X=\frac14\pm\epsilon/(2D)$
are replaced by $\frac14$; with the factor $\bar g=\tau$ this costs at most
$4\|z\|/n$. In the second line we apply both replacements. Dropping the
Fourier terms costs at most $16\|z\|/n$ for each of the three terms. The
error in the means is at most $(1+\tau^{-1})\delta_n\|z\|\le5\delta_n\|z\|$
for $h=\bar fz$ and at most $(1+\tau)\delta_n\|z\|\le2\delta_n\|z\|$ for
$h=\bar gz$; with the prefactors, the three terms contribute at most
$20$, $32$ and $20$, in units of $\|z\|/n$. Let $D_\tau$ be the operator
obtained from~\eqref{eq:full-linear} by these replacements.
By~\eqref{eq:model-T} and elementary algebra,
\begin{align*}
 (D_\tau z)_e&=\Bigl(\frac\tau2+\frac1\tau-1\Bigr)z_e
 +\Bigl(\frac\tau2-\frac1\tau\Bigr)\bar z_e-\frac\tau2\,\bar z_o,\\
 (D_\tau z)_o&=\frac14\Bigl(\tau+\frac1\tau\Bigr)z_o
 -\frac\tau2\,\bar z_e-\frac1{2\tau}\,\bar z_o,
\end{align*}
where the subscripts $e,o$ indicate even and odd weights. For $\tau=1$ this
is $D_1z=\frac12z-\frac12(\bar z_e+\bar z_o)\mathbf1$. For $\tau=1/3$,
writing $z=W_ez+W_oz+Cz$,
\[
 D_{1/3}z=\tfrac{13}6W_ez+\tfrac56W_oz-\tfrac12Cz
 -\tfrac16(\bar z_e+\bar z_o)\mathbf1.
\]
Since $\Pi_n$ annihilates constants and fixes $W_ez$, $W_oz$ and $Cz$, we
get $\Pi_nD_1=L_1$ and $\Pi_nD_{1/3}=L_J$ on $E_n$, whatever the actual
sizes of the two parity classes. For $\tau\in[1/4,1/2]$, the derivatives
with respect to $\tau$ of the three coefficients in each row of $D_\tau$
have absolute values adding up to at most $34$, so replacing $\tau_n$ by
$1/3$ costs at most $34|\tau_n-1/3|\,\|z\|\le34\|z\|/n$.

\emph{Deletion of the local directions.} The input pairs of a weight-$w$
label with an entry of weight one are counted by $N_w(1,w)=N_w(w,1)=2w$.
In the full-group row they contribute $-4w\tau/q_w^2$, so
\begin{equation}\label{eq:local-deletion}
 r_w=r_w^G+\frac{4w\tau}{q_w^2}.
\end{equation}
The logarithmic derivative of $4w\tau/(Xq_w^2)$ has norm at most
$32n/X\le2/n$. Collecting the errors, before projection they add up to at
most $(4+48+72+34+2)\|z\|/n=160\|z\|/n$, and since $\|\Pi_n\|\le2$,
\[
 \|D\Phi(0)-L\|\le\frac{320}{n}<\frac{512}{n}.
\]

\emph{The residual.} The counts $X/4$ and $c_\pm$ show that on the full
group the even and odd Ricci eigenvalues of the reference metric are
\[
 X\Bigl(1-\frac\tau2\Bigr),\qquad
 X\left[\left(\tfrac14+\tfrac{\epsilon}{2D}\right)\tau
       +\left(\tfrac14-\tfrac{\epsilon}{2D}\right)\tau^{-1}\right],
\]
and these agree for $\tau=1$ and $\tau=\tau_n$; this is the Einstein
condition for the two group metrics. Hence, by~\eqref{eq:local-deletion},
$\Phi(0)$ is the projection of the vector $(4w\tau/\bar q_w^{\,2})_w/X$,
whose entries are at most $16n/X$, and $\|\Phi(0)\|\le32n/X$.

\emph{Second derivatives.} In the coordinates $z$, every term
of~\eqref{eq:ricci} divided by $X$ is a constant multiple of $e^{c\cdot z}$
with $\|c\|_1\le3$, and the second derivative of $e^{c\cdot z}$ has norm
at most $\|c\|_1^2e^{c\cdot z}$. On the ball $\|z\|\le1/10$, the source term
$1/q_w$ contributes at most $4e^{1/10}$. The bracket terms have reference
values $\bar q_t/(\bar q_w\bar q_s)$ and $\bar q_w/(\bar q_s\bar q_t)$ at
most $16$, and the weights $N_w(s,t)/X$ add up to at most $1/2$ in each of
the two sums, which carry the factors $2$ and $1$. Including the
projection,
\[
 \|D^2\Phi(z)\|\le2\left(4e^{1/10}+9\cdot16e^{3/10}\cdot\tfrac32\right)
 =2\left(4e^{1/10}+216e^{3/10}\right)<1024.
\]
\end{proof}

\begin{theorem}\label{thm:local}
Let $n\ge2^{16}$. For each reference profile there is exactly one Einstein
solution $\bar qe^z$ with $z\in E_n$ and $\|z\|_\infty\le2^{-20}$, and it
satisfies $\|z\|_\infty\le256n/4^n$. The two solutions obtained in this way
are not proportional.
\end{theorem}
\begin{proof}
Put $r_0=2^{-20}$ and $\Psi(z)=z-L^{-1}\Phi(z)$ on $E_n$; the fixed points
of $\Psi$ are the zeros of $\Phi$. For $\|z\|\le r_0$,
Lemma~\ref{lem:local-linear} gives
\[
 \|D\Psi(z)\|\le\|L^{-1}\|\bigl(\|L-D\Phi(0)\|+\|D\Phi(0)-D\Phi(z)\|\bigr)
 \le6\Bigl(\frac{512}n+1024r_0\Bigr)<\frac14,
\]
and $\|\Psi(0)\|\le6\|\Phi(0)\|\le192n/4^n<r_0/2$; the last inequality
holds at $n=2^{16}$ because $4^n>384\cdot2^{20}n$, and $4^n/n$ increases.
Hence $\Psi$ maps the closed ball of radius $r_0$ into itself and is a
contraction there, so it has exactly one fixed point $z$ in this ball.
Moreover
\[
 \|z\|\le\frac{\|\Psi(0)\|}{1-1/4}\le8\|\Phi(0)\|\le\frac{256n}X.
\]
The ratio $q_3/q_2$ is invariant under scaling. For the solution near
$\bar q^{\,+}$ it lies in $[e^{-2r_0},e^{2r_0}]$, and for the solution
near $\bar q^{\,-}$ it is at most $\tau_ne^{2r_0}\le\frac12e^{2r_0}
<e^{-2r_0}$. The two solutions are therefore not proportional.
\end{proof}

\subsection{Exhaustion, stability and geometric distinction}\label{sec:classification}
\begin{proof}[Proof of Theorem~\ref{thm:classification}]
\emph{Exhaustion.} Let $n_j\to\infty$, and let $q^{(j)}$ be Einstein
solutions on $M_{n_j}$, normalized by $q_2=1$. By
Theorem~\ref{thm:apriori},
\[
 \frac{\log K(q^{(j)})}{n_j}
 \le\frac{83\log2}{n_j}+\frac{4608}{\sqrt{n_j}}
 \longrightarrow0,
\]
so by Theorem~\ref{thm:subexp} the maximum distance from $q^{(j)}$ to the
nearer of the two profiles that equal $1$ on even weights and $1$ or $1/3$
on odd weights tends to zero. The coefficients of these profiles are
bounded away from zero, so the same holds for the logarithms. Passing
to fixed volume subtracts the weighted mean of the logarithms, which at
most doubles the error, and replacing $1/3$ by $\tau_n$ changes the profile
by $O(2^{-n})$. Hence, for large $j$, $q^{(j)}$ lies, up to scaling, in one
of the two balls of Theorem~\ref{thm:local}.

Suppose there were arbitrarily large $n$ with an Einstein solution
different, up to scaling, from the two solutions of
Theorem~\ref{thm:local}. Choosing one for each such $n$ gives a sequence as
above, and for large $n$ its member lies in one of the balls of
Theorem~\ref{thm:local}; by local uniqueness it coincides with the solution
constructed there, a contradiction. Therefore, for all large $n$, there are at
most two solutions of~\eqref{intro:equations}, and
Theorem~\ref{thm:local} provides two for every $n\ge2^{16}$. Their
Einstein constants are positive, as for every Einstein solution
(Section~\ref{sec:geometry}).

\emph{Profiles.} Theorem~\ref{thm:local} gives a displacement $O(n4^{-n})$
in fixed-volume logarithmic coordinates. Normalizing by $q_2=1$ subtracts
the second coordinate from every logarithm, which at most doubles this
bound. This proves~\eqref{intro:profiles}.

\emph{Hessians.} Give $E_n$ the inner product $\langle\cdot,\cdot\rangle_m$.
The projections $W_e$, $W_o$ and $C$ are orthogonal for it, and their
ranks on $E_n$ are the number of even weights minus one, the number of odd
weights minus one, and one. Thus $L_1$ is positive definite, whereas $L_J$
has one negative eigenvalue (on the range of $C$) and $n-3$ positive ones.
By~\eqref{eq:variation}, along fixed-volume logarithmic coordinates
$d\Scal=-X\langle\Phi(z),dz\rangle_m$, and therefore
\begin{equation}\label{eq:hessian-final}
 \operatorname{Hess}_{z}\Scal(u,v)
 =-X\langle D\Phi(z)u,v\rangle_m,
 \qquad u,v\in E_n.
\end{equation}
In particular $D\Phi(z)$ is self-adjoint for $\langle\cdot,\cdot\rangle_m$.
At the solution $z_*$ near either reference, Lemma~\ref{lem:local-linear}
gives
\[
 \|D\Phi(z_*)-L\|_\infty
 \le\frac{512}{n}+1024\|z_*\|_\infty\longrightarrow0.
\]
The segment from $L$ to $D\Phi(z_*)$ consists of self-adjoint operators,
and they are all invertible for large $n$, because
$\|L^{-1}\|_\infty\|D\Phi(z_*)-L\|_\infty<1$. Hence $D\Phi(z_*)$ has the
same inertia as $L$, and~\eqref{eq:hessian-final} shows that the Hessian of
the scalar curvature is nondegenerate, negative definite at $q^+$ and with
exactly one positive eigenvalue at $q^-$.

\emph{Isometry classes.} By Corollary~\ref{cor:geometric} below, whose
proof uses only the parts of the theorem established above, no positive
multiple of $g_{q^+}$ is isometric to $g_{q^-}$.
\end{proof}

For an Einstein metric $g$ on $M_n$ with Einstein constant $\lambda(g)$, put
\[
 \mathscr I_n(g)=\frac{\lambda(g)}{4^n}
       \left(\frac{\Vol(g)}{V_{0,n}}\right)^{2/d}.
\]
Multiplying $g$ by $c>0$ multiplies $\Vol(g)$ by $c^{d/2}$ and $\lambda(g)$
by $c^{-1}$, so $\mathscr I_n$ is invariant under scaling; it is also
invariant under isometries.

\begin{corollary}\label{cor:geometric}
For all sufficiently large $n$, the Einstein constants of the two metrics
of Theorem~\ref{thm:classification}, normalized by $q_2=1$, satisfy
\begin{align}
 \lambda_n^+/4^n&=\tfrac12+O(n4^{-n}),\label{eq:lambda-plus}\\
 \lambda_n^-/4^n&=1-\tfrac12\tau_n+O(n4^{-n}),\label{eq:lambda-minus}
\end{align}
and
\begin{equation}\label{intro:invariants}
 \mathscr I_n(g_{q^+})\longrightarrow\frac12,\qquad
 \mathscr I_n(g_{q^-})\longrightarrow\frac5{6\sqrt3}.
\end{equation}
In particular, no positive multiple of $g_{q^+}$ is isometric to $g_{q^-}$.
\end{corollary}
\begin{proof}
On a fixed logarithmic neighborhood of either reference profile, the
row-sum norm of $D(r/X)$ is bounded uniformly in $n$: each term
of~\eqref{eq:ricci} is a multiple of $e^{c\cdot z}$ with $\|c\|_1\le3$, the
reference coefficients lie between $1/4$ and $1$, and the weights
$N_w(s,t)/X$ add up to at most $1/2$. Therefore the
displacement~\eqref{intro:profiles} changes $r/X$ by $O(n/X)$. At the
references, $r_w/X=\frac12+4w/X$ for $\bar q^{\,+}$, and
$r_w/X=1-\tau_n/2+O(n/X)$ for $\bar q^{\,-}$ by~\eqref{eq:local-deletion}.
This proves~\eqref{eq:lambda-plus} and~\eqref{eq:lambda-minus}.

By~\eqref{eq:volume},
\[
 \mathscr I_n(g_q)=\frac{\lambda(q)}X
 \exp\left(\frac1d\sum_{w=2}^nm_w\log q_w\right).
\]
For $q^+$ the exponent tends to zero. For $q^-$ it equals
$\frac kd\log\tau_n+O(n/X)$, where $k/d\to1/2$ and $\tau_n\to1/3$.
Consequently $\mathscr I_n(g_{q^+})\to\frac12$ and
$\mathscr I_n(g_{q^-})\to\frac56\cdot\frac1{\sqrt3}$. The two limits are
different because $25\ne27$, so $\mathscr I_n(g_{q^+})\ne
\mathscr I_n(g_{q^-})$ for large $n$, and no rescaling of $g_{q^+}$ is
isometric to $g_{q^-}$.
\end{proof}

\section{Control distances and circuit size}\label{sec:control}
\subsection{Quotient distance as a control cost}
For a traceless Hermitian Hamiltonian $A$, let $A_S$ be its component with
tensor support $S$, so that $iA_S\in W_S$, and put
\[
 \|A\|_0^2=D^{-1}\Tr(A^2),\qquad
 F_q(A)^2=\sum_{|S|\ge2}q_{|S|}\|A_S\|_0^2.
\]
An absolutely continuous path of unitaries with $\dot U=-iA(t)U$ almost
everywhere has control length $\int_0^1F_q(A(t))\,dt$. We work
projectively, so multiples of the identity in $A$ play no role, and local
terms cost nothing. For $h_1,h_2\in H_n$,
\[
 \mathcal C_q(h_1Uh_2)=\mathcal C_q(U),
\]
because left multiplication by $h_1$ does not change the coset $H_nU$, and
right multiplication by $h_2$ is an isometry of $M_n$ that fixes the base
point $H_n$. By right invariance, the distance between two arbitrary points
is $d_{g_q}(H_nU,H_nV)=\mathcal C_q(VU^{-1})$.

\begin{proposition}\label{qc:control}
For every positive coefficient vector $q$,
\begin{equation}\label{qc:operational}
 \mathcal C_q(U)=\inf\left\{
 \int_0^1F_q(A(t))\,dt:
 U(0)=I,\ U(1)\in H_nU,\ \dot U=-iA(t)U
 \right\}.
\end{equation}
If $b=\max_wq_w$ and $c_0=\pi/\sqrt3$, then
\begin{equation}\label{qc:diameter}
 \diam(M_n,g_q)\le c_0\sqrt b.
\end{equation}
\end{proposition}
\begin{proof}
Extend $g_q$ to a right-invariant metric on $G_n$ by giving the local Lie
algebra the inner product $Q$. Since $g_q$ is
$\operatorname{Ad}(H_n)$-invariant, this metric is also invariant under
left multiplication by $H_n$, and the projection $G_n\to M_n$ is a
Riemannian submersion~\cite{ONeill}. At $U$, the vertical vectors are the
$-iA_{\mathrm{loc}}U$ with $A_{\mathrm{loc}}$ local, and the horizontal
vectors are the $-iAU$ with $A$ supported on sets of at least two qubits.
Hence a path with $\dot U=-iA(t)U$ projects to a path of length
$\int F_q(A(t))\,dt$ in $M_n$, which proves that the infimum is at least
$\mathcal C_q(U)$. Conversely, a minimizing geodesic from $H_n$ to $H_nU$
has a horizontal lift starting at $I$; it has the same length and ends
in $H_nU$. This proves~\eqref{qc:operational}.

Let $e^{i\phi_1},\ldots,e^{i\phi_D}$ be the eigenvalues of a unitary
representative of a point of $\PU(D)$. Averaging over a global phase,
\[
 \frac1{2\pi}\int_0^{2\pi}\frac1D\sum_{j=1}^D
 \Arg(e^{i(\phi_j+\theta)})^2\,d\theta
 =\frac1{2\pi}\int_{-\pi}^{\pi}t^2\,dt=\frac{\pi^2}{3}.
\]
Hence for some $\theta$ the Hermitian matrix with eigenvalues
$\Arg(e^{i(\phi_j+\theta)})$ and the eigenvectors of the representative,
which is a logarithm of $e^{i\theta}$ times the representative, has
$\|\cdot\|_0^2\le\pi^2/3$. Removing its trace part does not change the
projective class of its exponential and does not increase $\|\cdot\|_0$.
The resulting constant Hamiltonian $A$ steers $I$ to the given point along
a path of length $F_q(A)\le\sqrt b\|A\|_0\le c_0\sqrt b$. By transitivity,
the same bound holds for any two points of $M_n$.
\end{proof}

Combined with the classification, Proposition~\ref{qc:control} bounds the
diameters of all invariant Einstein metrics.

\begin{corollary}\label{qc:all-Einstein}
For all sufficiently large $n$, every invariant Einstein metric on
$M_n$, normalized by $q_2=1$, satisfies
\[
 \diam(M_n,g_q)\le\frac\pi{\sqrt3}
 \bigl(1+O(n4^{-n})\bigr),
\]
where the implied constant does not depend on the metric.
\end{corollary}
\begin{proof}
By Theorem~\ref{thm:classification}, the metric is $g_{q^+}$ or
$g_{q^-}$. Both reference profiles have largest coefficient one,
so~\eqref{intro:profiles} gives $\max_wq_w=1+O(n4^{-n})$.
Apply~\eqref{qc:diameter}.
\end{proof}

In this cost model every Hamiltonian is available, at a price given by its
support-weighted norm; the difficulty of producing a many-body Hamiltonian
from two-body interactions is not part of the cost. This is why bounded
Einstein distances are compatible with the exponential gate counts of
Section~\ref{qc:circuits}.

\subsection{The normal distance and the two Einstein metrics}
For every invariant metric, the image of Haar measure on $G_n$ is the
normalized Riemannian volume of $M_n$, since both are probability measures
invariant under the transitive right action. In this subsection $U$ is Haar
distributed. Let
\[
 f_D(U)=d_{\PU(D),Q}(I,U)
\]
be the distance from the identity in $\PU(D)$ with the bi-invariant metric
$Q$.

\begin{lemma}\label{qc:group-concentration}
For every $D\ge2$,
\begin{equation}\label{qc:group-mean}
 f_D\le c_0,\qquad \E f_D\ge c_0-\frac{12}{c_0D},
\end{equation}
and for $t>0$,
\begin{equation}\label{qc:group-tail}
 \Pp[f_D\le\E f_D-t]\le\exp(-D^2t^2/4).
\end{equation}
\end{lemma}
\begin{proof}
The distance $f_D(U)$ is the smallest norm $\|X\|_0$ of a Hermitian $X$
with $e^{iX}=e^{i\theta}U$ for some $\theta$. Such an $X$ commutes with $U$,
so its eigenvalues are $\phi_j+\theta+2\pi k_j$ with integers $k_j$. For
fixed $\theta$ the best choice of the $k_j$ gives the principal arguments,
and for fixed lifts the best $\theta$ subtracts their mean; performing the
two minimizations in either order gives
\begin{equation}\label{qc:phaseformula}
 f_D(U)^2=\min_{\theta\in\mathbb R}\frac1D
 \sum_{j=1}^D\Arg(e^{i(\phi_j+\theta)})^2.
\end{equation}
In particular $f_D\le c_0$ by the phase average in the proof of
Proposition~\ref{qc:control}.

For Haar distributed $U$,
\begin{equation}\label{qc:trace-moment}
 \E|\Tr U^k|^2=\min(k,D),\qquad k\ge1,
\end{equation}
see~\cite{DS}. This also follows from the eigenvalue kernel
$K_D(\alpha,\beta)=(2\pi)^{-1}\sum_{r=0}^{D-1}e^{ir(\alpha-\beta)}$: the
diagonal contributes $D$ to $\E|\Tr U^k|^2$, and the two-point correction
is $-\max(D-k,0)$, the number of pairs $r,s\in\{0,\ldots,D-1\}$ with
$s-r=k$. The Fourier series
\[
 t^2=\frac{\pi^2}{3}+4\sum_{k\ge1}\frac{(-1)^k\cos(kt)}{k^2},
 \qquad -\pi\le t\le\pi,
\]
converges absolutely, and substituting $t=\Arg(e^{i(\phi_j+\theta)})$ and
summing over $j$ gives
\[
 \E\sup_\theta\left|
 \frac1D\sum_j\Arg(e^{i(\phi_j+\theta)})^2-c_0^2
 \right|
 \le\frac4D\sum_{k\ge1}\frac{\E|\Tr U^k|}{k^2}
 \le\frac4D\sum_{k\ge1}\frac{\sqrt k}{k^2}
 \le\frac{12}{D},
\]
since $\sum_kk^{-3/2}\le3$. By~\eqref{qc:phaseformula},
$\E(c_0^2-f_D^2)\le12/D$, and $c_0^2-f_D^2\ge c_0(c_0-f_D)$
gives~\eqref{qc:group-mean}.

The function $f_D$ is $1$-Lipschitz. For normalized Haar measure $\mu$, write
\[
 \Ent_\mu(F)=\int F\log F\,d\mu-\left(\int F\,d\mu\right)\log\left(\int F\,d\mu\right).
\]
The bi-invariant metric $Q$ on $\PU(D)$ has Ricci tensor $(D^2/2)Q$, so the
Bakry--\'Emery criterion~\cite{BGL} gives the logarithmic Sobolev inequality
\[
 \Ent_\mu(h^2)\le\frac4{D^2}\int|\nabla h|^2\,d\mu.
\]
Applied to $h=e^{sf_D/2}$, this yields a differential inequality for
$s\mapsto\log\E e^{sf_D}$ whose integration (the Herbst argument) gives
$\log\E e^{s(f_D-\E f_D)}\le s^2/D^2$ for all real $s$; since $f_D$ is only
Lipschitz, one applies this to smooth approximations. The exponential
Markov inequality with $s=-tD^2/2$ gives~\eqref{qc:group-tail}.
\end{proof}

\begin{theorem}\label{qc:typical}
Let $D=2^n$, and let $\mathcal C$ be $\mathcal C_0$ or $\mathcal C_{q^+}$.
Then $\mathcal C(U)\to c_0$ in probability and in every moment of fixed
positive order, and the diameter of the corresponding metric tends to
$c_0$. For each $\varepsilon>0$ there is $c_\varepsilon>0$ such that
$\Pp[\mathcal C(U)\le c_0-\varepsilon]\le e^{-c_\varepsilon4^n}$ for all
large $n$. Moreover
\begin{equation}\label{qc:Jensen-distance}
 \Pp[\mathcal C_{q^-}(U)<\pi/3-\varepsilon]
 \le e^{-c'_\varepsilon4^n},
 \qquad
 \limsup_{n\to\infty}\diam(M_n,g_{q^-})\le c_0.
\end{equation}
\end{theorem}
\begin{proof}
The normal metric induces on the identity component $\PU(2)^n$ of $H_n$
the product of the metrics of the factors. Each factor has a $\delta$-net
of size at most $(C/\delta)^3$, with an absolute constant $C$. Taking
$\delta/\sqrt n$-nets in all factors and translating them by the $n!$
qubit permutations gives a $\delta$-net of $H_n$, for the distance of
$G_n$, of size at most
\begin{equation}\label{qc:net}
 n!\,(C\sqrt n/\delta)^{3n}.
\end{equation}
The normal quotient distance is
$\mathcal C_0(U)=\min_{h\in H_n}d_{G_n,Q}(h,U)$.
Fix $0<\varepsilon<c_0$ and let $\delta=\varepsilon/4$. For large $n$,
Lemma~\ref{qc:group-concentration} gives $\E f_D>c_0-\varepsilon/4$. If
$\mathcal C_0(U)\le c_0-\varepsilon$, then some point of the net lies
within $c_0-3\varepsilon/4$ of $U$. For each fixed net point $h$, the
distance $d_{G_n,Q}(h,U)$ has the same distribution as $f_D(U)$, by Haar
invariance, so~\eqref{qc:group-tail} and a union bound give
\[
 \Pp[\mathcal C_0\le c_0-\varepsilon]
 \le n!\,(4C\sqrt n/\varepsilon)^{3n}
       e^{-4^n\varepsilon^2/16}
 \le e^{-4^n\varepsilon^2/32}
\]
for all large $n$. Together with $\mathcal C_0\le c_0$, this proves the
assertions for $\mathcal C_0$.

By~\eqref{intro:profiles}, the quadratic forms of $g_{q^+}$ and $g_0$
differ by a factor $e^{\pm O(n4^{-n})}$, so their distance functions differ
by a factor $1+O(n4^{-n})$. This transfers the convergence in probability
and the lower-tail estimate to $\mathcal C_{q^+}$. The distances are
uniformly bounded, so all moments of fixed order converge, and the diameter
lies between the mean distance and an upper bound that tends to $c_0$. For
$q^-$, the smallest coefficient tends to $1/3$ and the largest to one, so
$g_0/(3+o(1))\le g_{q^-}\le(1+o(1))g_0$; since $c_0/\sqrt3=\pi/3$, the
same comparison proves~\eqref{qc:Jensen-distance}.
\end{proof}

Corollary~\ref{qc:all-Einstein} and Theorem~\ref{qc:typical} prove
Theorem~\ref{intro:control}.

\subsection{Fixed-accuracy two-body circuit size}\label{qc:circuits}
Consider circuits made of arbitrary two-qubit gates on arbitrary pairs of
qubits and arbitrary single-qubit gates, without ancillas, where only the
two-qubit gates are counted. Let $G_{2,\varepsilon}(U)$ be the least number
of two-qubit gates in a circuit $V$ with
\[
 d_{\mathrm{op}}(U,V):=
 \min_{\theta\in\mathbb R}\|U-e^{i\theta}V\|_{\mathrm{op}}
 \le\varepsilon.
\]
The circuit thus has to approximate $U$ itself, up to a global phase, and
not merely the coset $H_nU$.

\begin{proposition}\label{qc:gates}
If $q_2=1$ and $b=\max_wq_w$, then for $0\le\varepsilon<2$,
\begin{equation}\label{qc:gatebound}
 G_{2,\varepsilon}(U)\ge
 \frac{\mathcal C_q(U)-2\sqrt b\arcsin(\varepsilon/2)}{c_0}.
\end{equation}
\end{proposition}
\begin{proof}
By the phase average in the proof of Proposition~\ref{qc:control} with
$D=4$, every two-qubit gate has a projective Hermitian logarithm $A$ with
$\|A\|_0\le c_0$, and embedding into the $n$-qubit system preserves this
normalized norm. The only nonlocal part of $A$ has support of size two, so
$F_q(A)\le\|A\|_0\le c_0$ because $q_2=1$. Concatenating these paths with
the free paths of the single-qubit gates gives $\mathcal C_q(V)\le c_0G_2$
for a circuit $V$ with $G_2$ two-qubit gates. If
$d_{\mathrm{op}}(U,V)\le\varepsilon$, then $e^{-i\theta}UV^{-1}$ has, for a
suitable $\theta$, eigenvalues $e^{i\psi}$ with
$|\psi|\le2\arcsin(\varepsilon/2)$, and its principal logarithm gives a
path from $H_nV$ to $H_nU$ of cost at most $2\sqrt b\arcsin(\varepsilon/2)$.
The triangle inequality and minimization over circuits prove the claim.
\end{proof}

No bound in the opposite direction holds for the Einstein costs, even at
fixed accuracy. The following counting argument is a version of the
classical circuit counting of~\cite{Knill,Shende}, with the gate set and
the operator-norm accuracy made explicit.

\begin{proposition}\label{qc:approx-hard}
Fix $0<\varepsilon<1/2$. There are constants $a_\varepsilon,b_\varepsilon>0$
such that, for all sufficiently large $n$,
\begin{equation}\label{qc:approx-hard-bound}
 \Pp\left[G_{2,\varepsilon}(U)\le
                  a_\varepsilon\frac{4^n}{n}\right]
 \le e^{-b_\varepsilon4^n}.
\end{equation}
\end{proposition}
\begin{proof}
A circuit with $j$ two-qubit gates can be rewritten as $j$ two-qubit gates
followed by at most $n$ single-qubit gates: move each single-qubit gate
past the two-qubit gates that do not touch its qubit and absorb it into
the next one that does; the gates that are never absorbed remain at the
end.

Fix the sequence of qubit pairs on which the two-qubit gates act; there are
at most $n^{2j}$ such sequences. Cover $\SU(4)$ and $\SU(2)$ by
operator-norm $\delta$-nets of sizes at most $(C/\delta)^{15}$ and
$(C/\delta)^3$; the scalar phases of individual gates can be ignored
projectively. With $\delta=\varepsilon/(J+n)$, a telescoping estimate shows
that a product of at most $J+n$ unitaries changes by at most $\varepsilon$
in operator norm when each factor is replaced by a net point, so the
circuits with at most $J$ two-qubit gates have an $\varepsilon$-net of
size at most
\begin{equation}\label{qc:circuit-net}
 (J+1)n^{2J}
 \left(\frac{C(J+n)}{\varepsilon}\right)^{15J+3n}.
\end{equation}
A target within $\varepsilon$ of such a circuit is within $2\varepsilon$ of
a net point. A ball of radius $2\varepsilon$ in the projective operator
norm is contained in a ball of radius $2\arcsin\varepsilon<\pi/3$ for the
normal metric, and Lemma~\ref{qc:group-concentration} bounds the Haar
probability of the latter by $e^{-b_04^n}$, where $b_0>0$ depends only on
$\varepsilon$.

Let $J=\lfloor a_\varepsilon4^n/n\rfloor$. For large $n$,
$\log(C(J+n)/\varepsilon)\le C_\varepsilon n$, so the logarithm
of~\eqref{qc:circuit-net} is at most
$C'_{\varepsilon}a_\varepsilon4^n+O_\varepsilon(n^2)$. Choose
$a_\varepsilon$ so that the first term is at most $b_04^n/4$. A union bound
then proves~\eqref{qc:approx-hard-bound} with $b_\varepsilon=b_0/2$.
\end{proof}

By Corollary~\ref{qc:all-Einstein}, the Einstein distances are bounded
uniformly in the target and in large $n$, so any fixed polynomial in $n$
and in such a distance is bounded by a polynomial in $n$. By
Proposition~\ref{qc:approx-hard}, most targets need at least
$a_\varepsilon4^n/n$ two-qubit gates. Therefore no fixed polynomial in $n$ and
$\mathcal C_q(U)$ bounds the circuit size of every target. Thus~\eqref{qc:gatebound} turns a large geometric cost into a lower bound
on gates, but a small cost in the Einstein metrics does not produce a short
circuit; this is consistent with the dependence on penalties found
in~\cite{NDGD,BrownEquivalence}.

\section{Locality penalties and the two-body limit}\label{sec:penalty}
\subsection{The exact interval of nonnegative Ricci curvature}
For $T\ge1$, let $g_T$ be the metric with coefficients
\begin{equation}\label{qc:two-level}
 q_2=1,\qquad q_w=T\quad(3\le w\le n).
\end{equation}
Two-body directions keep unit cost, while directions of weight at least
three have length penalty $\sqrt T$. In this section we put
\[
 a_2=9\binom n2,\quad A_n=48n-80,\quad b_n=24(n-2),
 \quad h_n=\frac X2-24n+40,
\]
\[
 k_w=2w(3n-2w-1),\qquad j_w=12\cdot\mathbf1_{\{w=3\}}.
\]

\begin{lemma}\label{qc:ricci-window}
For $n\ge4$, the Ricci eigenvalues of~\eqref{qc:two-level} are
\begin{equation}\label{qc:two-level-ricci}
\begin{aligned}
 r_2(T)&=A_n-b_nT+h_n/T^2,\\
 r_w(T)&=\frac{X/2+4w}{T}+j_w(T-1/T)
       +2(k_w-j_w)(1/T-1/T^2),\qquad w\ge3.
\end{aligned}
\end{equation}
There is a unique $T_R>1$ such that
\begin{equation}\label{qc:TR}
 b_nT_R^3-A_nT_R^2-h_n=0.
\end{equation}
For $T\ge1$, $\Ric(g_T)\ge0$ exactly when $T\le T_R$, and
$\Ric(g_T)>0$ exactly when $T<T_R$. None of these metrics is Einstein.
Moreover
\begin{equation}\label{qc:TR-asymptotic}
 T_R\sim\left(\frac{4^n}{48n}\right)^{1/3}.
\end{equation}
\end{lemma}
\begin{proof}
For a label of weight $w$, an input pair whose first entry has weight two
has second entry of weight $w+1$ or $w-1$: the weight-two label meets the
support of the given label in one site, where the letters differ, or in two
sites, where they differ at exactly one. By~\eqref{eq:counts},
\[
 N_w(2,w+1)=6w(n-w),\qquad
 N_w(2,w-1)=2w(w-1),
\]
and these add up to $k_w$. For $w\ge3$ there are therefore $2k_w-j_w$
input pairs with an entry of weight two; $j_w$ of them have two such
entries, which happens only for $w=3$, where $N_3(2,2)=12$. For $w=2$, the
horizontal input pairs with an entry of weight two are those with entries
of weights $(2,3)$ or $(3,2)$, $24(n-2)$ in all, because
$N_2(2,3)=12(n-2)$ and pairs with an entry of weight one are excluded.
Substituting these counts and the total $X/2-4w$ from~\eqref{eq:balance}
into~\eqref{eq:energy} gives~\eqref{qc:two-level-ricci}.

We have $h_n>0$, $b_n>0$ and $k_w\ge j_w$. For $w\ge3$ every term
in~\eqref{qc:two-level-ricci} is nonnegative when $T\ge1$, and the first
one is positive. On the other hand,
\[
 r_2(1)=X/2+8>0,\qquad
 r_2'(T)=-b_n-2h_n/T^3<0,\qquad r_2(T)\to-\infty,
\]
so $r_2$ has exactly one zero $T_R>1$; it solves~\eqref{qc:TR}, and
$r_2(T)\ge0$ exactly for $T\le T_R$, with strict inequality for $T<T_R$.
This proves the assertions about the Ricci curvature. To see that no $g_T$
is Einstein, put $u=T-1\ge0$; subtracting the two formulas gives
\begin{equation}\label{qc:not-Einstein}
 \begin{split}
 T^2(r_3-r_2)={}&4+(X/2+12n-56)u\\
 &+(24n-28)u^2+(24n-36)u^3>0.
 \end{split}
\end{equation}
Finally, put $t_n=(h_n/b_n)^{1/3}$ and $z=T_R/t_n$. Dividing~\eqref{qc:TR}
by $b_nt_n^3=h_n$ gives $z^3-\alpha_nz^2=1$ with
$\alpha_n=A_n/(b_nt_n)=O(t_n^{-1})$. The function $z^3-\alpha_nz^2-1$ is
negative at $z=1$ and positive at $z=1+\alpha_n$, so $1<z<1+\alpha_n$ and
$T_R=t_n+O(1)$. Since $t_n^3=h_n/b_n\sim X/(48n)$, this
gives~\eqref{qc:TR-asymptotic}.
\end{proof}

\subsection{Volume, typical distance and gate lower bounds}
Let $\omega_d$ be the volume of the Euclidean unit ball in dimension $d$,
and put
\[
 \vrad(g)=\left(\frac{\Vol(g)}{\omega_d}\right)^{1/d}.
\]
If $\Ric(g)\ge0$, Bishop's inequality bounds the volume of a ball of
radius $r$ by $\omega_dr^d$, so a ball of radius $s\,\vrad(g)$ contains at
most the fraction $s^d$ of the total volume. We compute the volume radius
of $g_0$ and of $g_T$.

\begin{lemma}\label{qc:volume}
For $N\ge2$, let $\Gamma$ denote Euler's gamma function and put
\[
 v(N)=\frac{(2\pi)^{N(N+1)/2-1}N^{-N^2/2}}
              {\prod_{j=1}^N\Gamma(j)}.
\]
Then $v(2)=\pi^2$, and the normal volume and volume radius of $M_n$ are
\begin{equation}\label{qc:normal-volume}
 V_{0,n}=\frac{v(2^n)}{n!\,\pi^{2n}},\qquad
 R_{0,n}:=\vrad(g_0)\longrightarrow e^{1/4}.
\end{equation}
For the penalty metrics,
\begin{equation}\label{qc:penalty-volume}
 R_n(T):=\vrad(g_T)
 =R_{0,n}T^{(d-a_2)/(2d)},\qquad
 \frac{R_n(T)}{\sqrt T}\longrightarrow e^{1/4}
\end{equation}
uniformly for $1\le T\le T_R$.
\end{lemma}
\begin{proof}
We first compute the volume of $U(N)$ for the metric $-\Tr$, in agreement
with~\cite{Macdonald}. The orbit of the first coordinate vector $e_1$ is
$S^{2N-1}$, with stabilizer $U(N-1)$. The skew-Hermitian matrices
orthogonal to $\mathfrak u(N-1)$ map $e_1$ to vectors $(ia,z)$ with
$a\in\mathbb R$ and $z\in\mathbb C^{N-1}$, and have squared norm
$a^2+2|z|^2$ for $-\Tr$. Compared with the round metric of the unit sphere,
the quotient metric is therefore larger by the factor $\sqrt2$ in $2N-2$
real directions, and its volume is $2^{N-1}\Vol(S^{2N-1})$. The fibers are
isometric to $(U(N-1),-\Tr)$, so
\[
 \frac{\Vol(U(N),-\Tr)}{\Vol(U(N-1),-\Tr)}
 =2^{N-1}\Vol(S^{2N-1})=\frac{(2\pi)^N}{(N-1)!}.
\]
Starting from $\Vol(U(1))=2\pi$, induction gives
\[
 \Vol(U(N),-\Tr)=
 \frac{(2\pi)^{N(N+1)/2}}{\prod_{j=1}^N\Gamma(j)}.
\]
The center $\{e^{i\theta}I\}$ is a circle of length $2\pi\sqrt N$ for
$-\Tr$. Dividing by this length gives the volume of $\PU(N)$ for the
quotient metric, and rescaling the metric by $N^{-1}$ multiplies the volume
by $N^{-(N^2-1)/2}$; the result is $v(N)$, the volume of $\PU(N)$ for
$-N^{-1}\Tr$. The identity component $\PU(2)^n$ of $H_n$, with the metric
induced by $Q$, is a product of $n$ copies of $(\PU(2),-\frac12\Tr)$, of
volume $v(2)^n=\pi^{2n}$, and $H_n$ has $n!$ components. Since
$G_n\to M_n$ is a Riemannian submersion for the normal metric with fibers
isometric to $H_n$, this proves the first formula
in~\eqref{qc:normal-volume}.

Summing Stirling's formula gives
\[
 \sum_{j=1}^D\log\Gamma(j)
 =\tfrac12D^2\log D-\tfrac34D^2+O(D\log D),
\]
and hence
\[
 \log v(D)=\frac X2\log(2\pi/X)+\frac{3X}{4}+O(D\log D).
\]
Also $\log\omega_d=\frac d2\log(2\pi e/d)+O(\log d)$. Since $d=X-1-3n$ and
$\log(n!)=O(n\log n)$, dividing $\log(V_{0,n}/\omega_d)$ by $d$ shows that
$\log R_{0,n}\to\frac34-\frac12=\frac14$.

There are $a_2$ two-body directions. Multiplying the metric on the other
$d-a_2$ directions by $T$ multiplies the volume by $T^{(d-a_2)/2}$, which
gives the formula for $R_n(T)$. Finally, $a_2/d=O(n^2/4^n)$ and
$\log T_R=O(n)$, so $T^{-a_2/(2d)}\to1$ uniformly for $1\le T\le T_R$.
\end{proof}

For comparison, the Einstein metrics satisfy
\[
 \vrad(g_{q^+})\longrightarrow e^{1/4},\qquad
 \vrad(g_{q^-})\longrightarrow e^{1/4}3^{-1/4},
\]
by~\eqref{eq:volume} and~\eqref{intro:profiles}, because the odd labels
make up asymptotically half of the horizontal labels. Thus their volume
radii, like their diameters, stay bounded in two-body units.

\begin{theorem}\label{qc:large-distance}
For $n\ge4$, $1\le T\le T_R$ and $0<s<1$,
\begin{equation}\label{qc:tail}
 \Pp[\mathcal C_T(U)\le sR_n(T)]\le s^d,
 \qquad \mathcal C_T(U)\le c_0\sqrt T.
\end{equation}
Uniformly over all sequences $T=T(n)$ in this interval, the median and the
mean of $\mathcal C_T(U)$ and the diameter of $g_T$ are $\Theta(\sqrt T)$,
and $\E\mathcal C_T^r=\Theta(T^{r/2})$ for each fixed $r>0$. The choice
\begin{equation}\label{qc:explicit-penalty}
 T_n=\left(\frac{4^n}{192n}\right)^{1/3},\qquad n\ge5,
\end{equation}
gives positive Ricci curvature and distance scale
$\Theta(2^{n/3}n^{-1/6})$.
\end{theorem}
\begin{proof}
By Lemma~\ref{qc:ricci-window}, $\Ric(g_T)\ge0$, so Bishop's comparison
theorem~\cite{Besse} gives $\Vol B(x,r)\le\omega_dr^d$ for every ball.
Dividing by the total volume and using Haar invariance gives the first
inequality in~\eqref{qc:tail}; the second is Proposition~\ref{qc:control}
with $b=T$.

By Lemma~\ref{qc:volume}, $sR_n(T)\ge s(e^{1/4}-o(1))\sqrt T$ uniformly for
$1\le T\le T_R$. Hence, for fixed $s\in(0,1)$, $\mathcal C_T(U)$ is at least
a fixed positive multiple of $\sqrt T$ with probability at least $1-s^d$,
and it is never larger than $c_0\sqrt T$. This gives all the stated
orders: for instance,
$\E\mathcal C_T^r\ge(sR_n(T))^r(1-s^d)$, the median is at least $sR_n(T)$
once $s^d<1/2$, and the diameter lies between the mean and $c_0\sqrt T$.

For~\eqref{qc:explicit-penalty}, note that $T_n\ge1$, $h_n\ge X/4$ and
$b_n\le24n$ when $n\ge5$. Since $X/(4T_n^2)=48nT_n$,
\[
 r_2(T_n)\ge A_n-24nT_n+\frac{X}{4T_n^2}
 =A_n+24nT_n>0,
\]
and the other Ricci eigenvalues are positive by
Lemma~\ref{qc:ricci-window}. The distance scale is
$\sqrt{T_n}=(4^n/(192n))^{1/6}$.
\end{proof}

Lemma~\ref{qc:ricci-window}, Theorem~\ref{qc:large-distance} and
Theorem~\ref{qc:penalty-limit} below prove Theorem~\ref{intro:penalty}.
Within the interval of nonnegative Ricci curvature, the largest typical
distance obtained in this way is of order $\sqrt{T_R}\asymp(4^n/n)^{1/6}$.
The volume argument follows Brown~\cite{BrownVolume}.

Combining Proposition~\ref{qc:gates} with~\eqref{qc:tail} shows that, with
probability at least $1-s^d$,
\begin{equation}\label{qc:penalty-gates}
 G_{2,\varepsilon}(U)\ge
 \frac{sR_n(T)-2\sqrt T\arcsin(\varepsilon/2)}{c_0}.
\end{equation}
For $s=1/2$ and $\varepsilon=1/10$, the right side is at least a positive
constant times $\sqrt T$ for large $n$, since $R_n(T)/\sqrt T\to e^{1/4}$
and $2\arcsin(1/20)<0.11$. This bound on the number of gates comes from
curvature and volume alone. It is much weaker than the counting bound of
Proposition~\ref{qc:approx-hard}, which gives order $4^n/n$ for most
targets.

\subsection{Strictly two-body motion}
Let $\mathcal D_2$ be the $G_n$-invariant distribution on $M_n$ whose fiber
at the base point $H_n$ is $\mathfrak m_2$, with the inner product
$Q|_{\mathfrak m_2}$, and let $d_{\mathrm{cc}}$ be the associated
sub-Riemannian distance. Its bracket flag is
\[
 \mathcal D_2^{(1)}=\mathcal D_2,\qquad
 \mathcal D_2^{(r+1)}=\mathcal D_2^{(r)}
       +[\mathcal D_2,\mathcal D_2^{(r)}],
\]
where the brackets are taken between local smooth sections. Local controls
remain free. Whenever $q_2=1$, the restriction of $g_q$ to $\mathcal D_2$
is this inner product, whatever the higher coefficients are.

\begin{theorem}\label{qc:flag}
The distribution $\mathcal D_2$ is bracket generating and equiregular of
step $n-1$, and its bracket flag at the base point is
\begin{equation}\label{qc:flagformula}
 (\mathcal D_2^{(r)})_{H_n}
 =\bigoplus_{2\le w\le\min(r+1,n)}\mathfrak m_w,
 \qquad 1\le r\le n-1.
\end{equation}
Its Hausdorff dimension is
\begin{equation}\label{qc:Hdim}
 \mathsf Q_2=\sum_{w=2}^n(w-1)3^w\binom nw
 =\left(\frac{3n}{4}-1\right)4^n+1.
\end{equation}
In particular, $d_{\mathrm{cc}}$ is finite and continuous, and the
restrictions of the two Einstein metrics of
Theorem~\ref{thm:classification} to $\mathcal D_2$ define the same
distance $d_{\mathrm{cc}}$.
\end{theorem}
\begin{proof}
The preimage of $\mathcal D_2$ in $\mathfrak{su}(D)$ is the sum of the local
algebra and the modules $W_S$ with $|S|=2$, and the flag at the base point
is obtained from iterated brackets in $\mathfrak{su}(D)$ modulo the local
algebra. A two-site support that is disjoint from another support commutes
with it; otherwise the union of the two supports has at most one more
site. Brackets with local directions preserve supports. Consequently the brackets
of order $r$ lie in the sum of the $W_S$ with $|S|\le r+1$, which is the
inclusion $\subseteq$ in~\eqref{qc:flagformula}.

Conversely, suppose that $W_S$ has been obtained, let $j\notin S$, and
choose $i\in S$. If $P_a\in W_S$ and $P_b\in W_{\{i,j\}}$ have different
letters at $i$, then $[P_a,P_b]$ is, up to a scalar, the tensor product of
the commutator of the two letters at $i$, the letter of $b$ at $j$, and the
letters of $a$ at the other sites of $S$. The commutators at $i$ span
$\mathfrak{su}(2)$ and the other factors can be chosen freely, so these
brackets span $W_{S\cup\{j\}}$. Induction on $|S|$ gives equality
in~\eqref{qc:flagformula}. The full support first appears at step $n-1$,
and since $G_n$ acts transitively, the ranks of the flag are the same at
every point. The same mechanism of support growth appears in the study of
many-body controllability~\cite{Bremner}.

By the Chow--Rashevskii theorem, $d_{\mathrm{cc}}$ is finite and continuous.
For an equiregular distribution, the ball-box theorem gives the Hausdorff
dimension as the sum over the layers of the flag of the layer dimension
times its step~\cite{Mitchell,ABB}. Here $W_S$ first occurs at step
$|S|-1$, so $\mathsf Q_2=\sum_{w\ge2}(w-1)m_w$. The identities
\[
 \sum_{w=0}^n3^w\binom nw=4^n,\qquad
 \sum_{w=0}^nw3^w\binom nw=3n4^{n-1}
\]
give $\sum_{w=0}^n(w-1)3^w\binom nw=(3n/4-1)4^n$. The term with $w=0$ equals
$-1$ and the term with $w=1$ vanishes, which proves~\eqref{qc:Hdim}.
Finally, all metrics with $q_2=1$ assign the same length to every path
tangent to $\mathcal D_2$, so their restrictions define the same distance.
\end{proof}

\begin{theorem}\label{qc:penalty-limit}
For each fixed $n\ge3$ there is a constant $C_n$ such that
\begin{equation}\label{qc:penalty-rate}
 0\le d_{\mathrm{cc}}(x,y)-d_{g_T}(x,y)
 \le C_n T^{-1/[2(n-1)]}
\end{equation}
for all $T\ge1$ and all $x,y\in M_n$.
Consequently $d_{g_T}$ increases uniformly to $d_{\mathrm{cc}}$ as
$T\to\infty$.
\end{theorem}
\begin{proof}
Since $g_T$ increases with $T$, so does $d_{g_T}$. Every path tangent to
$\mathcal D_2$ has the same length for all $T$, so
$d_{g_T}\le d_{\mathrm{cc}}$. By homogeneity it suffices to bound
$d_{\mathrm{cc}}(H_n,H_nU)$. Lift a minimizing $g_T$-geodesic from $H_n$
to $H_nU$ horizontally to a path of unitaries starting at $I$, of length
$\ell=d_{g_T}(H_n,H_nU)$ and driven by $A=A_2+A_{>2}$, where $A_2$ is the
two-body part and $A_{>2}$ the part supported on at least three qubits.
Let $V$ be the endpoint of the path driven by $A_2$ alone. Variation of
constants and unitarity give
\begin{equation}\label{qc:pruning}
 \|U-V\|_{\mathrm{op}}
 \le\int_0^1\|A_{>2}(t)\|_{\mathrm{op}}\,dt
 \le\sqrt{D/T}\,\ell,
\end{equation}
because $\|A\|_{\mathrm{op}}\le\sqrt D\|A\|_0$ and
$\sqrt T\|A_{>2}\|_0\le F_T(A)$. The path driven by $A_2$ is tangent to
$\mathcal D_2$ and has length at most $\ell$, so
$d_{\mathrm{cc}}(H_n,H_nV)\le\ell$.

For a bracket-generating distribution of step $n-1$, the ball-box theorem
gives $d_{\mathrm{cc}}(x,y)\le C_nd_{g_0}(x,y)^{1/(n-1)}$ for nearby
points, and by compactness for all $x,y$ after enlarging $C_n$. A principal
logarithm of $UV^{-1}$ gives $d_{g_0}(H_nV,H_nU)\le C_n'\|U-V\|_{\mathrm{op}}$
when the right side is small. Since
$\ell\le\diam(M_n,d_{\mathrm{cc}})<\infty$,~\eqref{qc:pruning} and the
triangle inequality give, for large $T$,
\[
 d_{\mathrm{cc}}(H_n,H_nU)
 \le d_{\mathrm{cc}}(H_n,H_nV)+d_{\mathrm{cc}}(H_nV,H_nU)
 \le\ell+C_nT^{-1/[2(n-1)]}.
\]
The remaining bounded range of $T$ is covered by enlarging $C_n$.
\end{proof}

The constant $C_n$ obtained in this way depends on $n$ through the ball-box
theorem, so Theorem~\ref{qc:penalty-limit} describes the limit
$T\to\infty$ for a fixed number of qubits.

\end{document}